\documentclass[12pt]{article}

\AtBeginDocument{
  \addtocontents{toc}{\normalsize}
}
\pdfoutput=1

\usepackage{amsmath,amssymb,amsfonts,amsthm,amscd,mathrsfs}
\usepackage{xcolor}
\definecolor{darkblue}{rgb}{0.1,0.1,.7}
\usepackage[]{version}
\usepackage{float}
\usepackage[]{graphicx}
\usepackage[]{latexsym}
\usepackage{geometry}
\usepackage{subcaption}
\usepackage[utf8]{inputenc}
\usepackage[all,cmtip]{xy}

\usepackage{ifthen}
\usepackage{tikz}
\usepackage{array,setspace,yfonts,dsfont,bbm,colonequals}
\usepackage[colorlinks, linkcolor=darkblue, citecolor=darkblue, urlcolor=darkblue, linktocpage]{hyperref} 

\usepackage{cite}
\usepackage{xspace}
\usepackage{empheq}
\usepackage{xcolor}

\numberwithin{equation}{section}
\newcommand{\ba}{\begin{equation}\begin{aligned}}
\newcommand{\ea}{\end{aligned}\end{equation}}

\newcommand{\cO}{\mathcal O}

\newcommand{\reef}[1]{(\ref{#1})}
\newcommand{\be}{\begin{equation}}
\newcommand{\ee}{\end{equation}}

\newcommand{\bea}{\begin{equation}\begin{aligned}}
\newcommand{\eea}{\end{aligned}\end{equation}}

\newcommand{\ud}{\mathrm d}

\newcommand{\Df}{{\Delta_\phi}}

\newcommand{\calC}{\mathcal{C}}

\begin{document}

\vspace*{-.6in} \thispagestyle{empty}
\begin{flushright}
\end{flushright}
\vspace{1cm} {\Large
\begin{center}
{\bf A functional approach to the numerical conformal bootstrap}\\
\end{center}}
\vspace{1cm}
\begin{center}
{Miguel F.~Paulos${}^{1}$, Bernardo Zan${}^{2,1}$}\\[1cm] 
{
\small
${}^1$ {\em Laboratoire de Physique de l'\'Ecole Normale Sup\'erieure\\ PSL University, CNRS, Sorbonne Universit\'es, UPMC Univ. Paris 06\\ 24 rue Lhomond, 75231 Paris Cedex 05, France
}\\
${}^2$ {\em Theoretical Particle Physics Laboratory (LPTP), Institute of Physics,\\EPFL, Lausanne, Switzerland }
\normalsize
}
\\
\end{center}

\begin{center}
	{\texttt{miguel.paulos@lpt.ens.fr, bernardo.zan@epfl.ch} 
	}
	\\
\end{center}

\vspace{4mm}

\begin{abstract}
We apply recently constructed functional bases to the numerical conformal bootstrap for 1D CFTs. We argue and show that numerical results in this basis converge much faster than the traditional derivative basis. In particular, truncations of the crossing equation with even a handful of components can lead to extremely accurate results, in opposition to hundreds of components in the usual approach. We explain how this is a consequence of the functional basis correctly capturing the asymptotics of bound-saturating extremal solutions to crossing. We discuss how these methods can and should be implemented in higher dimensional applications.

\end{abstract}
\vspace{2in}


\newpage

{
\setlength{\parskip}{0.05in}
\tableofcontents
\renewcommand{\baselinestretch}{1.0}\normalsize
}


\setlength{\parskip}{0.1in}
\newpage

\section{Introduction}\label{sec:introduction}
More than ten years ago, \cite{Rattazzi:2008pe} introduced a simple and effective method to extract information from crossing equations in conformal field theories. In its simplest guise,\footnote{In this example the equation expresses crossing symmetry for a correlator of four identical fields $\phi$ on a line, when decomposed into $SL(2,\mathbb R)$ blocks.} the equation takes the form
%
\bea
\sum_{\cO\in \phi\times \phi} \lambda_{\phi \phi\cO}^2 F_{\Delta_{\cO}}(z|\Df)=0.\label{eq:crossingintro},
\eea%
with kinematically determined functions $F_{\Delta}$ given below. This equation should be thought of as expressing an infinite set of sum rules, labeled by the cross-ratio $z$, on the OPE data $\lambda_{\phi \phi \cO}$ between primary operators $\phi, \phi, \cO$. The basic idea of \cite{Rattazzi:2008pe} is as follows.\footnote{For a recent review and an extensive list of references see \cite{Poland:2018epd}.} Firstly we abstract from any particular CFT realization and think of the above as a general constraint on all possible consistent OPEs. Secondly we act on the crossing equation with suitably chosen linear functionals $\omega$ to get simpler sum rules:
\bea
\sum_{\Delta\in S} a_\Delta \omega(\Delta|\Df)=0, \qquad \omega[F_{\Delta}(z|\Df)]\equiv \omega(\Delta|\Df)\,,
\eea
where for unitary solutions we have $a_{\Delta_{\cO}}\equiv\lambda_{\phi\phi\cO}^2\geq 0$. We can now make  various assumptions on the set $S$ of allowed quantum numbers, on the external dimension $\Df$, or even the allowed values of $a_\Delta$ for specific $\Delta$. Depending on these assumptions, it may be possible to find $\omega(\Delta|\Df)$ non-negative in $S$. The sum rule is then impossible to satisfy for positive $a_{\Delta}$, which means our original assumptions cannot hold for any unitary CFT. In this way, it is possible to derive bounds on CFT data such as scaling dimensions and OPE coefficients.

In practice, the search for such functionals is done numerically inside a finite dimensional linear vector space. This amounts to truncating the crossing constraints to a more manageable finite subset. We want the size of this space to be as large as possible to get the strongest bounds, but we are constrained in doing this by the available computational resources. Clearly we would like to choose this space in a way which gives us the most stringent constraints possible for a fixed dimensionality. In the original reference \cite{Rattazzi:2008pe} this choice corresponds to taking $\omega$ to be a finite sum of derivatives with respect to the cross-ratio $z$ at a specific point. For its simplicity and effectiveness, this has remained the default choice up to now.\footnote{See \cite{CastedoEcheverri2016} for some alternative explorations.}

Very recently, new bases of functionals were constructed in \cite{Mazac2019a} (see also \cite{Kaviraj2018,Mazac:2016qev,MazacOPE2018,Mazac:2018,Mazac:2018biw} for related works) and found to lead to exact, optimal, bounds on CFT data. The goal of this note is to explore how these bases can be used to obtain better numerical bounds. As we will see, one of the key points is that in these bases each individual functional automatically encodes the expected asymptotics and spectrum of bound-saturating (extremal \cite{El-Showk:2016mxr,ElShowk:2012hu}) solutions to crossing. This leads to a decoupling between dynamics at small and large scaling dimensions, which means the truncated bootstrap equations not only fully capture the interesting physics, they may also be effectively completed to the full set of constraints once the truncation size is large enough.

In this note we will consider two simple applications to test our proposal. In the first we will bootstrap the generalized free boson solution using a functional basis associated to free fermions. This somewhat perverse choice was chosen as a worse-case scenario for our approach, and yet we will still find that it is a significant improvement over the derivative basis. For our second application we will find a family of universal bounds on OPE coefficients in 1D. The bound is saturated by a family of solutions to crossing which neatly interpolates between the generalized free boson and fermionic solutions. Near the boson point it is described by (boundary correlators of) a free scalar in AdS$_2$ with a $\phi^4$ interaction. We are able to easily find the approximate exact bound, obtaining several digits accuracy using a handful of components. 

The outline of this note is as follows. In the next section we review the functional bases obtained in \cite{Mazac2019a}. Section \ref{sec:reviewbounds} reviews the link between bounds and extremal solutions, and explains why the new basis is expected to lead to highly improved numerical bounds. In section \ref{sec:applications} we do the two numerical applications explained above, which confirm these expectations. We conclude with an outlook on higher dimensional applications.

This note is complemented by two appendices. The reader interested in using the functionals will find the explicit form of the associated integral kernels in appendix \ref{app:funcbasis}. Furthermore for integer or half-integer external dimensions, the kernels simplify and we explain how to compute the functional actions explicitly in appendix \ref{app:computation}.

\section{Setup}
\subsection{Crossing symmetry}
We are interested in studying the crossing equation for 1D CFT correlators of identical operators $\phi$:
\bea
\langle \phi(x_1)\phi(x_2)\phi(x_3)\phi(x_4)\rangle=\frac{\mathcal G(z)}{x_{13}^{2\Df}x_{24}^{2\Df}},\qquad x_{ij}=x_i-x_j,\quad z=\frac{x_{12}x_{34}}{x_{13}x_{24}}\,.
\eea
For $z\in (0,1)$ the correlator can be expressed in terms of $SL(2,\mathbb R)$ blocks as,
\bea
\mathcal G(z)=\sum_{\cO\in \phi\times \phi} a_{\Delta_\cO} \frac{G_{\Delta_{\cO}}(z)}{z^{2\Df}},\qquad a_{\Delta_{\cO}}=\lambda_{\phi\phi \cO}^2\,,
\eea
with the conformal block $G_{\Delta}(z)$ given by
\bea
G_{\Delta}(z):=z^{\Delta}\,_2F_1(\Delta,\Delta,2\Delta,z).
\eea
Crossing symmetry is the statement $\mathcal G(z)=\mathcal G(1-z)$. For a generic CFT this leads to the equation
\bea
\sum_{\Delta} a_{\Delta} F_{\Delta}(z|\Df)=0,
\eea
with $F_{\Delta}(z|\Df)=z^{-2\Df} G_{\Delta}(z)-(z\leftrightarrow 1-z)$. We will often simplify notation by omitting the dependence on $\Df$. For unitary CFTs the sum ranges over all scaling dimensions $\Delta\geq 0$.
We will study this equation as a mathematical set of constraints on the $a_{\Delta}$ which hold for any and all (unitary) CFTs.

Two simple and important solutions to these constraints are associated with generalized free fields in $D=1$, which describe boundary correlators of free fields in AdS$_2$. Introduce:
\bea
a_{\Delta}^{\text{free}}=\frac{2\,\Gamma(\Delta)^2\Gamma(\Delta+2\Df-1)}{\Gamma(2\Df)^2\Gamma(2\Delta-1)\Gamma(\Delta-2\Df+1)}\,.
\eea
Then we have
\bea
F_0(z|\Df)+\sum_{n=0}^{+\infty} a_{\Delta_n^{B,F}}^{\text{free}}\, F_{\Delta_n^{B,F}}(z|\Df)=0\label{eq:gffsol}
\eea
with $\Delta_n^B=2\Df+2n$ and $\Delta_n^F=1+2\Df+2n$. For $z\in (0,1)$ the corresponding correlators are
\bea
\mathcal G^{B,F}(z)=\pm 1+z^{-2\Df}+(1-z)^{-2\Df}
\eea
with the $+$ and $-$ signs for (B)osons and (F)ermions respectively.

\subsection{Free functional bases}

We will now review the analytic functionals constructed in \cite{Mazac2019a}. There are two functional bases which are in a sense dual to the simple solutions to crossing of the previous subsection. We will denote a generic functional by $\omega$ and functional actions will be abbreviated as
\bea
\omega(\Delta)\equiv \omega\left[F_{\Delta}(z|\Df)\right].
\eea

Each basis is made up of infinite sets of functionals $\alpha_n, \beta_n$ with $n$ a non-negative integer, and satisfying the duality conditions
\bea
\alpha_n^F(\Delta_m^F)&=\delta_{nm},&\qquad \partial_{\Delta} \alpha_n^F(\Delta_m^F)&=0,\\
\beta_n^F(\Delta_m^F)&=0,&\qquad \partial_{\Delta} \beta_n^F(\Delta_m^F)&=\delta_{nm}\,, \label{eq:dualityfermion}
\eea
for the fermionic (F) basis, and
\bea
\alpha_n^B(\Delta_m^B)&=\delta_{nm},&\qquad \partial_{\Delta} \alpha_n^B(\Delta_m^B)&=-\delta_{0m}c_n, \\
\beta_n^B(\Delta_m^B)&=0,&\qquad \partial_{\Delta} \beta_n^B(\Delta_m^B)&=\delta_{nm}-\delta_{0m}d_n\,,\label{eq:dualityboson}
\eea
for the bosonic (B) basis, with $\beta_0^B\equiv 0$. The coefficients appearing here are given by
\bea
c_n=\frac 12 \partial_n d_n, \qquad d_n=\frac{(\Df)_n^4(4\Df-1)_{2n}}{(n!)^2(2\Df)_n^2\,(4\Df+2n-1)_{2n}}.\label{eq:coeffs}
\eea
The fact that $c_n,d_n$ are non-zero is related to the fact the generalized free boson solution admits a relevant deformation that does not introduce new states. Thinking of the free boson as being the holographic dual to a massive scalar field $\phi$ in AdS$_2$, this deformation is accounted for by the $\phi^4$ interaction \cite{Mazac2019a}.

The action of the functionals is defined by the formula:
\bea
\omega[F_{\Delta}]=\frac 12 \int_{\frac 12}^{\frac 12+i\infty}\!\!\ud z f_{\omega}(z) F_\Delta(z)+\int_{\frac 12}^1 \ud z g_\omega (z) F_\Delta(z) \label{eq:contour}
\eea
with $g_\omega(z)=\pm (1-z)^{2\Df-2} f_{\omega}(\frac{1}{1-z})$ and the $+$ $(-)$ sign for boson (fermion) functionals respectively. The general functional kernels $f_\omega$ are given very explicitly in appendix \ref{app:funcbasis}. They are chosen such that not only the duality relations hold, but also that the functionals $\omega$ are compatible with the crossing equation, in the sense that:
\bea
F_0+\sum_{\Delta} a_{\Delta} F_{\Delta}=0\qquad \Rightarrow \qquad \omega\left[F_0+\sum_{\Delta} a_{\Delta} F_{\Delta}\right]=\omega(0)+\sum_{\Delta} a_{\Delta} \omega(\Delta).
\eea
The infinite sum appearing in the crossing equation means this swapping condition is not trivial, and it in fact constrains the behaviour of $f(z)$ near $z=0,1$ and $\infty$ \cite{Rychkov:2017tpc}. 

One of the main results of \cite{Mazac2019a} is that the crossing equation is equivalent to the set of sum rules arising from applying a complete functional basis:
\bea
\sum_{\Delta} a_{\Delta}F_{\Delta}(z)&=0\qquad \mbox{for all}\quad z\in (0,1)\nonumber\\&\qquad\mbox{is equivalent to}\nonumber\\[7pt]
  \sum_{\Delta} a_{\Delta} \omega(\Delta)&=0 \quad \mbox{for all}\quad \omega\in\left\{\alpha_n, \beta_n,\quad n\in \mathbb N_{\geq 0}\right\}.
\eea
The equivalence holds whether we use the bosonic or the fermionic basis of functionals. Because of this equivalence between the crossing equation and the sum rules involving the functional actions $\alpha_n(\Delta), \beta_n(\Delta)$ it is clearly of utmost importance to obtain expressions for these, or at least to have the means for their speedy numerical evaluation. In the derivative basis, this would be analogous to being able to compute arbitrary derivatives of conformal blocks efficiently. In our case, a useful result is that for each basis, for $\Delta>\Delta_m$ a contour rotation argument leads to the expression
\bea
\omega_m(\Delta)=[1\pm \cos\pi(\Delta-2\Df)]\int_0^1 \ud z\, g_{\omega_m}(z)\frac{G_{\Delta}(z)}{z^{2\Df}} \label{eq:funcaction}
\eea
where again the + ($-$) sign corresponds to the bosonic (fermionic) functionals. This expression is essentially what lies at the origin of the duality conditions, thanks to the oscillating prefactor. The limited validity of this formula comes about due to singularities in $g(z)$ as $z\to 0$. However, if one manages to evaluate the integral above exactly, we can of course extend the result to all $\Delta$ by analytic continuation. 

With the explicit forms of the $f_\omega(z)$ given in appendix \ref{app:funcbasis}, it is relatively straightforward to evaluate numerically the functional actions for any $\Delta$, using either \reef{eq:contour} or \reef{eq:funcaction}, and this is all we need for bootstrap applications.\footnote{In our experience, it is faster to numerically evaluate \eqref{eq:funcaction} when available, rather than \eqref{eq:contour}.} However, the kernels take on particularly simple expressions when $\Df$ is a half-integer, for the fermion basis, or an integer, for the bosonic basis. With these simplified kernels we are able compute the functional actions analytically, and this is  discussed in appendix \ref{app:computation}. These analytic expressions can be evaluated much more rapidly than by doing the integrals numerically. It would be very useful to be able to determine similar analytic results for any $\Df$, although this is strictly speaking not necessary for numerical applications.

\subsubsection*{Examples}

Since the discussion above is perhaps unfamiliar, it may be useful to consider a couple of concrete examples. For general $\Df$, $f(z)$ are sums of $_3F_2$ hypergeometric functions. In the special case $\Df=1$ we have instead, for instance, the simple expression:
\bea
f_{\alpha_0^B}(z)=\frac{2 \left((z-1) z+(z-2) (z+1) (2 z-1) \coth ^{-1}(1-2 z)+1\right)}{\pi ^2 (z-1) z}.
\eea
The functional action can now easily be computed numerically by plugging in this expression into \reef{eq:contour} with $g_{\alpha_0^B}(z)=(1-z)^{2\Df-2}f_{\alpha_0^B}\left(\frac 1{1-z}\right)$. Equivalently, we can use the methods described in appendix \ref{app:computation} to get an analytic expression. Either way the motivated reader can easily check that $\alpha_0^B(0)=-2$, $\alpha_0^B(2+2n)=\delta_{n,0}$, with the general shape of $\alpha_0^B(\Delta)$ as in figure \ref{fig:schem}. This is perfectly consistent with the fact that for $\Df=1$ the generalized free boson solution takes the form:
\bea
F_0(z)+2\, F_{2}(z)+\sum_{n\geq 1}^\infty a_n F_{2+2n}(z)=0.
\eea
Similarly, for $\Df=1/2$ the fermionic basis also simplifies. We have for instance:
\bea
g_{\beta_0^F}(z)=\frac{(1-z) \left(2 z^2+z+2\right)}{\pi ^2 z^2}\,.
\eea
Instead of using \reef{eq:contour}, in this case we will do the integral \eqref{eq:funcaction} directly, finding the analytic result\footnote{This particular functional and corresponding functional action first appeared in \cite{Mazac:2016qev}.}
\bea
\beta_0^F(\Delta)&=\frac{2 \sin^2\left(\frac{\pi\Delta}2\right)}{\pi^2}\frac{\Gamma(2\Delta)}{\Gamma(\Delta)^2}\,\bigg[\frac{1}{(\Delta -2) (\Delta +1)}+\\
&+\Gamma (\Delta )^3 \left(\, _3\tilde{F}_2(\Delta ,\Delta ,\Delta ;2 \Delta ,\Delta +1;1)-2 \Delta  \, _3\tilde{F}_2(\Delta ,\Delta ,\Delta +1;2 \Delta ,\Delta +2;1)\right)\bigg]=\\
&=\frac{2 \sin^2\left(\frac{\pi\Delta}2\right)}{\pi^2}\frac{\Gamma(2\Delta)}{\Gamma(\Delta)^2} \bigg\{\frac{1}{(\Delta -2) (\Delta +1)}+\\
&+\left(\Delta(\Delta -1)+\frac 12\right)\left[ \psi ^{(1)}\left(\frac{\Delta }{2}\right)-\psi ^{(1)}\left(\frac{\Delta +1}{2}\right)\right]-2\bigg\}
\eea
where $ _3 \tilde F_2$ stands for the regularized hypergeometric function. One can now explicitly check that the properties \eqref{eq:dualityfermion} are satisfied: the term in parenthesis is finite for every value of $\Delta=\Delta_m^F=2+2m$ for $m>0$, while it has a single pole for $\Delta=\Delta^F_0=2$. However, the prefactor has double zeroes at $\Delta=\Delta_m^F=2+2m$ for every $m\ge 0$, and we have indeed that $\beta_n^F(\Delta_m^F)=0$. 

\section{Bounds and the choice of basis}
\label{sec:reviewbounds}
The goal of this section is to review how bounds on CFT data are associated to extremal functionals and sparse solutions to crossing, and how this perspective explains why our proposed bases are ideally suited for numerics.
 In the next subsection we remind the reader of the basic idea of \cite{Rattazzi:2008pe}, and the link between optimal bounds and extremal solutions to the crossing equation \cite{El-Showk:2016mxr,ElShowk:2012hu}. In subsection \ref{sec:advantages} we will argue for our new basis of functionals. To do this we will use the link between extremal solutions and bounds, studying the former to understand the latter. In particular we will show that the functional bases described in the previous section strongly constrain the form of such solutions, leading to fast convergence of numerical bootstrap computations.

\subsection{Bounds and Extremality}
\label{sec:bounds}
Consider a set of solutions to crossing of the form
\bea
F_0(z)+a_{0} F_{\Delta_0}+\sum_{\Delta\geq \Delta_g} a_\Delta F_{\Delta}(z)=0\,.\label{eq:opecross}
\eea
We would like to determine an upper bound on $a_0$ \cite{Rattazzi2011}. To do this we apply a linear functional $\omega$ to this equation, setting without loss of generality $\omega(\Delta_0)=1$. We will consider $\omega$ to lie inside some linear vector space $\mathcal W_N$ of finite dimension. We get
\bea
a_{0}=-\omega(0)-\sum_{\Delta\geq \Delta_g} a_{\Delta} \omega(\Delta)\,.
\eea
Hence for any solution to the crossing equation \reef{eq:opecross} we have the bound
\bea
a_{0}\leq  -\omega^{\mbox{\tiny opt},N}(0)\equiv \underset{\omega\in \mathcal W_N}{\mbox{min}}\, -\!\omega(0)\quad\mbox{such that}\quad \omega(\Delta)\geq 0 \quad \mbox{for all}\quad \Delta\geq \Delta_g.
\eea
If $\hat \omega_n$ are a basis for $\mathcal W_N$, then we can define the dual space to be:
\bea
\mathcal W_N^*=\mbox{span}\left\{\vec F_{\Delta}(z), \quad \Delta\geq 0\right\},\qquad \vec F_{\Delta}:=\left(\hat\omega_1(\Delta),\ldots,\hat\omega_N(\Delta)\right).
\eea
Then we can also determine a bound by solving a problem in this space:
\bea
a_{0}\leq a^{\mbox{\tiny opt},N}_0\equiv 
\mbox{max}\, a_{0}\quad \mbox{such that}\quad \vec F_0+a_0 \vec F_{\Delta_0}+\sum_{\Delta\geq \Delta_g} a_{\Delta}\vec  F_{\Delta}=0\,,
\eea
for some $a_{\Delta}\geq 0$, since increasing $N$ adds more constraints and hence lowers $a_0$. These kinds of problems are known as semi-infinite linear programs \cite{Reemtsen1998}, and the theory of linear convex optimisation gives us an important result:
\bea
-\omega^{\mbox{\tiny opt},N}(0)=a_{0}^{\mbox{\tiny opt},N}.
\eea
In other words the functional problem and the dual problem are equivalent. 

\begin{figure}
	\centering
\includegraphics[width=8cm]{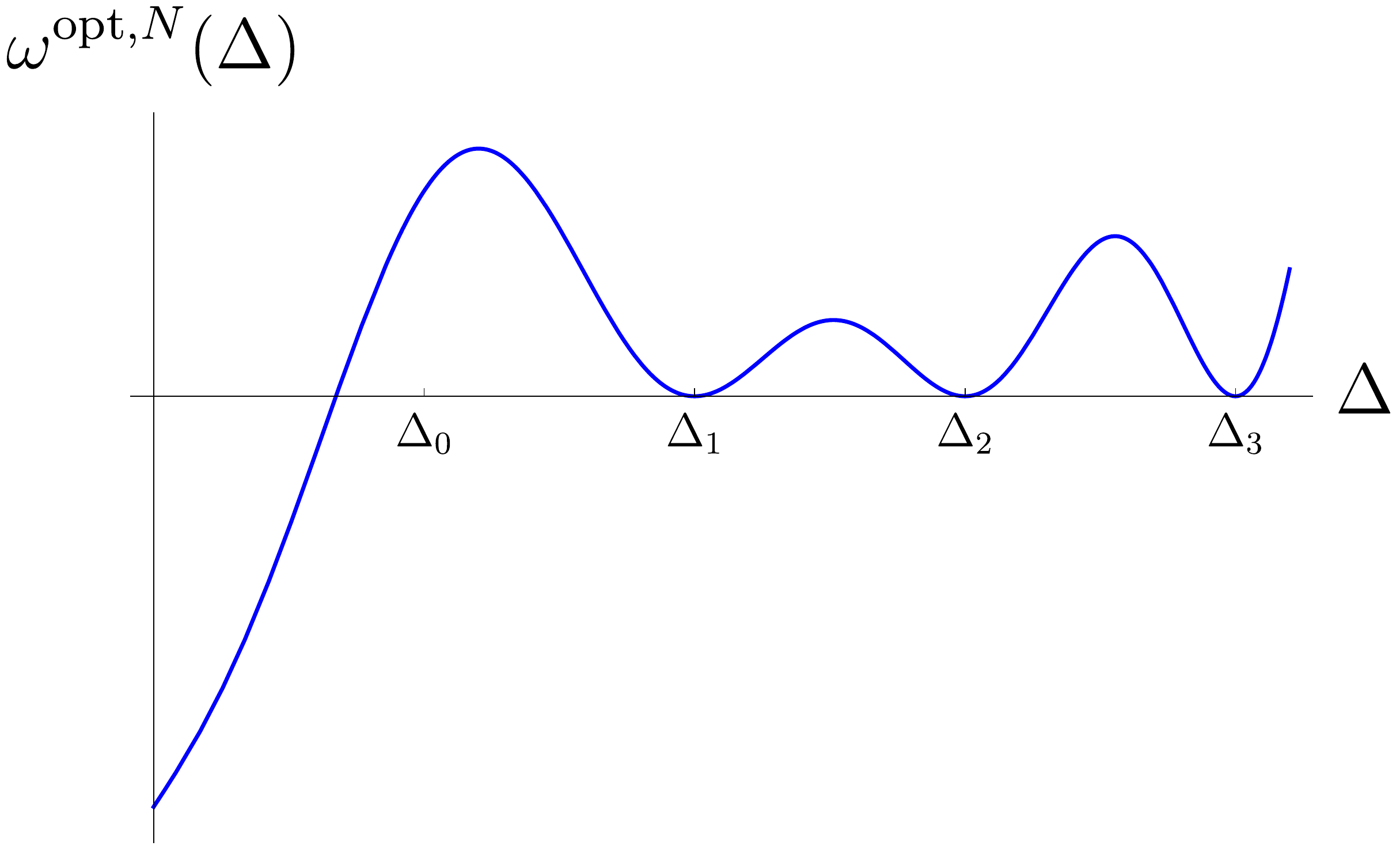}
\caption{Schematic form of the optimal functional. Such a functional provides a valid upper bound on the OPE coefficient at $\Delta_0$ for any choice of $\Delta_g$ down to where the functional first becomes negative.}
\label{fig:schem}
\end{figure}

This deserves some explanations. In the dual space formulation, we always find an explicit solution to the truncated form of the crossing equation (i.e. that involving the $\vec F_{\Delta}$), since we are satisfying the constraints throughout the maximization process. Meanwhile in the original formulation of the problem one constructs functionals which are supposed to rule out such solutions. How can these problems then be equivalent? The reason is that for any fixed $N$, at optimality there is indeed a solution to crossing but a very special one. The statement is that there exists a (generically unique) set of $a_n\geq 0$, $\Delta_n$ with $n=1,\ldots N-1$ such that not only
\bea
\vec F_0+a_{0}^{\mbox{\tiny opt},N} \vec F_{\Delta_0}+\sum_{n=1}^{N-1} a_{n}\vec  F_{\Delta_n}=0,\label{eq:funcconds1}
\eea
but furthermore
\bea
\omega^{\mbox{\tiny opt},N}(\Delta_n)=0,\qquad \partial_\Delta \omega^{\mbox{\tiny opt},N}(\Delta_n)=\delta_{\Delta_n,\Delta_g} \qquad \mbox{if} \quad a_n>0. \label{eq:funcconds2}
\eea
Note that these conditions are clearly necessary to make sure that $\omega(\Delta)$ is positive for $\Delta\geq\Delta_g$, see figure \ref{fig:schem}. The conditions \reef{eq:funcconds1} together \reef{eq:funcconds2} are the so-called Karush-Kuhn-Tucker second order optimality conditions. They are necessary conditions for optimality, with sufficiency being guaranteed as long as the functional $\omega$ is positive above $\Delta_g$.

For the kinds of problems we are interested in the conformal bootstrap, the fact that the set of allowed $\Delta$ is continuous typically leads to very sparse (extremal) solutions. For the 1D bootstrap problems considered in this note and elsewhere in the literature, one always finds (experimentally) that only $\lceil\frac{N-1}2\rceil$ operators appear in the extremal solution (beyond the identity and that at $\Delta_0$).\footnote{A similar statement apparently holds for the modular bootstrap when ignoring spin \cite{Afkhami-Jeddi:2019zci}.
The most general situation, which holds for instance in higher dimensions, is analysed in \cite{El-Showk:2016mxr}. }

Thus, extremal functionals go hand in hand with extremal solutions to crossing. What is perhaps less obvious is that at extremality we get not only the functional we were looking for, but actually a full basis for $\mathcal W_N$.
To see this very explicitly, let us consider the case where none of the $\Delta_n$ sit at the gap, set $N$ to be an odd number and consider the case where the extremal solution has only $(N-1)/2$ non-zero $a_n$. In this case the $N$ constraints from the crossing equation neatly match up with the unknown $(N+1)/2$ OPE coefficients (i.e the previously mentioned $(N-1)/2$ non-zero $a_n$ and $a_0$) and $(N-1)/2$ scaling dimensions of the extremal solution. We can then choose:
\bea
\mathcal W_N^*=\mbox{span}\left\{\vec F_{\Delta_0}; \vec F_{\Delta_n}, \partial_{\Delta} \vec F_{\Delta_n}, \quad n=1,\ldots, \frac{N-1}2\right\}.
\eea
We now construct a basis for the space of functionals $\mathcal W_N$ by setting
\bea
\mathcal W_N=\mbox{span}\left\{\alpha_0; \alpha_n, \beta_n, \quad n=1,\ldots,\frac{N-1}2\right\}
\eea
with functionals $\alpha,\beta$: satisfying the duality conditions:
\bea
\alpha_n(\Delta_m)&=\delta_{nm},&\qquad \partial_{\Delta}\alpha_n(\Delta_m)&=-\delta_{0m} C_n,&\qquad \\
\beta_n(\Delta_m)&=0,&\qquad \partial_{\Delta}\beta_n(\Delta_m)&=\delta_{nm}-\delta_{0m} D_n.\label{eq:orthoint}
\eea
with $\beta_0\equiv 0$ and some constants $C_n$, $D_n$. These constants, as well as the functionals, can be obtained completely explicitly by putting the basis vectors of $\mathcal W_N$ into a matrix and computing its inverse \cite{El-Showk:2016mxr}, leading to expressions in terms of the original set of $\hat \omega_i$. In particular, we see that our original OPE maximization functional $\omega^{\text{opt},N}$ is nothing but $\alpha_0$, as it satisfies \reef{eq:funcconds2}. Essentially the $\alpha$ functionals provide OPE bounds while $\beta$ functionals provide bounds on dimensions of operators, subject to certain assumptions. These bounds must exist, as they encode the uniqueness of the extremal solution under such assumptions.

To summarize, the message of this section is that optimal bounds are associated to extremely simple solutions to crossing containing as few operators as possible, which saturate these bounds. Such solutions in turn determine a pair of bases for $\mathcal W$ and $\mathcal W^*$. Hence, bounds on the CFT data follow from the study of these extremal solutions and their associated functionals.\footnote{Gliozzi \cite{Gliozzi2013} was the first to propose that sparseness (what we call extremality) can be used to bootstrap interesting solutions to crossing, even in the absence of unitarity or positivity.} 

\subsection{Arguments for the new bases}
\label{sec:advantages}

From now on we shift our perspective from determining bounds to constructing extremal solutions to the truncated crossing equations:
\bea
\vec F_{0}+a_{0} \vec F_{\Delta_0}+\sum_{m=1}^{\frac{N-1}2} a_{m} \vec F_{\Delta_m}&=0\\
\Leftrightarrow \hat \omega_n(0)+a_{0} \hat \omega_n(\Delta_0)+\sum_{m=1}^{\frac{N-1}2} a_{m} \hat \omega_n(\Delta_m)&=0, \qquad n=1,\ldots, N. \label{eq:truncated}
\eea
This is because as we have just reviewed, understanding and constructing these solutions automatically leads to the desired bounds. 

We are particularly interested in how the truncated solution approaches the exact one. As we dial up $N$, the number of operators in the solution increases, and the CFT data of those operators already in the solution changes, eventually approaching their exact values. Depending on the choice of basis elements $\hat \omega_1, \ldots \hat \omega_N$, the rate at which we approach the exact extremal solution, and with it the exact bounds, can be very different as we shall see. 

The original choice of \cite{Rattazzi:2008pe} set $\hat \omega_n=\partial_{z}^{2n-1}\big|_{z=\frac 12}$, and this has been the default approach ever since. We will argue that a better choice is to use the functionals $\alpha_n^{B,F},\beta_n^{B,F}$. Whether bosonic or fermionic functionals should be used depends on the expected asymptotics of the extremal solution. Although the full set of bosonic functionals is equivalent to the fermionic one, for finite $N$ we could choose a mixed basis combining elements from the two. For the sake of the arguments in this section we will stick to just using the bosonic basis, and set
\bea
\mathcal W_N=\text{span}\left\{\alpha_0^B; \alpha_n^B,\beta_n^B, n=1,\ldots \frac{N-1}2\right\}, \label{eq:basisfuncs}
\eea
with  $N$ odd.

To make our case we will focus on the OPE maximization problem of an operator with dimension $\Delta_0<2\Df+1$, the same problem which will be considered numerically in the next section. We will start by choosing $\Delta_0=2\Df$ and then discuss perturbations around this point up to second order. This problem was already considered and solved in \cite{Mazac2019a}, but here our focus will be on understanding the dependence of the solution on the truncation size. At the end we will give a more general argument that justifies why this basis should be generally useful for any bootstrap problem.

\subsubsection*{Perturbative arguments - zeroth order}

Consider first the case where $\Delta_0=2\Df$. We must search for a functional $\omega^{\text{opt},N}$ in $\mathcal W_N$ which will give us the best possible bound. In this case it is not hard to see that the optimal functional is $\alpha_0^B$, independently of $N$. Indeed it can be checked that $\alpha_0^B(\Delta)\geq 0$ for $\Delta\geq \Delta_0$ (and that $\alpha_0^B(\Delta_0)=1$). Hence $\alpha_0^B$ definitely provides a valid bound. That this bound is optimal can be checked by computing
\bea
-\alpha_0^B(0)=a_{\Delta_0^B}^{\text{free}},
\eea
which establishes optimality since the generalized free boson solution \reef{eq:gffsol} saturates it. This is not too surprising, since the exact functional bases were constructed precisely so as to be dual to the exact generalized free solutions. The extremal solution in this case
will consist of $\frac{N+1}2$ vectors with dimensions $\Delta_0^B,\ldots, \Delta_{\frac{N-1}2}^B$, with
\bea
\vec F_0+a_{\Delta_0^B}^{\text{free}} \vec F_{\Delta_0}+\sum_{n=1}^{\frac{N-1}2} a_{\Delta_n^B}^{\text{free}} \vec F_{\Delta_n^B}=0\,.
\eea
That this equation holds is a simple consequence of the duality conditions \reef{eq:dualityboson}. In other words, the truncated extremal solution is obtained by simply considering the first few operators in the exact generalized free boson solution, which appear with their {\em exact} scaling dimensions and OPE coefficients. As we add more functionals by increasing $N$ the solution to crossing systematically approaches the exact one, one operator at a time, in the best way possible. If we are only interested in obtaining the bound on $a_{\Delta_0}$, a truncation with $N=1$ will. Of course in the derivative basis the correct result would only be achieved with $N=\infty$. 

\subsubsection*{Perturbative arguments - first order}

For a more interesting result, consider instead maximizing the OPE coefficient of an operator whose dimension $\Delta_0$ is close to but not quite $2\Df$. We expect that the extremal solution (and accordingly, the optimal bound) can be obtained by perturbing around the generalized free boson. We will do our analysis by expanding in $g\equiv \Delta_0-2\Df$:
\bea
\Delta_m\sim \sum_{k=0}^{2 }g^{k}\Delta_m^{(k)}, \qquad a_{\Delta_m}\sim \sum_{k=0}^{2}g^{k}a_{m}^{(k)},
\eea
where the zeroth order values are the free ones. Expanding to leading order the truncated crossing equations \reef{eq:truncated} and using the duality relations \reef{eq:dualityboson} we find
\bea
a_{m}^{(1)}= a_{0}^{(0)} c_m, \qquad a_{m}^{(0)} \Delta_m^{(1)}=a_{0}^{(0)} d_m,
\eea
which fixes the extremal solution completely, including the new optimal bound
\bea
a_{0}^{\text{opt}}= a_0^{(0)}\left(1+g\, c_0\right)+O(g^2)\,.
\eea
These results are completely independent of the truncation size, and are therefore exact for any $N$. So, just as before, the only role that the truncation size has is to systematically add more operators which appear with their correct dimensions and OPE coefficients (up to $O(g^2)$ corrections). In particular, we again find that the exact OPE bound is captured, to leading order in $g$, even with $N=1$.

On the other hand, given this extremal solution to crossing we can also reconstruct the new extremal functionals.. Imposing the duality relations on the modified solution  we can find for instance\footnote{Note that in this process we are not free to choose the $C_n$, $D_n$, rather their precise values emerge automatically after imposing the other duality relations.}
\bea
\alpha_0=(1+g c_0)\left[\alpha_0^B-g\sum_{k=1}^{\frac{N-1}2} \Delta_k^{(1)} \partial^2_{\Delta}\alpha_0^B(\Delta_k^{(0)})\beta_k^{B}\right] +O(g^2)\,.
\eea
Notice that this result now depends on the explicit truncation, and hence this is actually not the correct exact functional for any finite $N$. The reason for this is that this functional still has double zeros at $\Delta_m=2\Df+2m$ for $m>N/2$ (i.e. above the truncation level), instead of the corrected $O(g)$ ones. In spite of this, this functional does compute the correct new bound to $O(g^2)$, $-\alpha_0(0)=a_{0}^{(0)}+g a_{0}^{(1)}$, independently of the truncation size.

\subsubsection*{Perturbative arguments - second order}

We now go to quadratic order in $g$. In this case the crossing equations lead to:
\bea
a_{m}^{(2)}&=a_0^{(1)} c_m-\frac 12 \sum_{n=0}^{\frac{N-1}2} a_n^{(0)} (\Delta_n^{(1)})^2 \partial_{\Delta}^2 \alpha^B_m(\Delta_n^{(0)})\,,\\
a_{m}^{(0)}\Delta_m^{(2)}+a_{m}^{(1)}\Delta_m^{(1)}&=a_0^{(1)}d_m-\frac 12 \sum_{n=0}^{\frac{N-1}2} a_n^{(0)} (\Delta_n^{(1)})^2 \partial_{\Delta}^2 \beta^B_m(\Delta_n^{(0)})\,.
\eea
The results now depend explicitly on $N$, a dependence which is directly inherited from that of the functionals at the previous order. To proceed we need to know something about the functional actions and the anomalous dimensions. Using the asymptotics derived in \cite{Mazac2019a} (quoted below in equation \reef{eq:asymptotic}) as well as the large $n$ behaviour of $d_n$ from \reef{eq:coeffs}, we have:
\bea
\Delta_n^{(1)}\underset{n\gg 1}{\sim} 1/n^2, \qquad  a_n^{(0)} \partial_{\Delta}^2 \omega^B_m(\Delta_n^{(0)})\underset{n\gg 1}{\sim}  1/n^{3}, 
\eea
with $\omega=\alpha,\beta$, and holding $m$ fixed in the last equality. We conclude that both scaling dimensions and OPE coefficients are expected to converge to their exact ($N=\infty$) values at a rate $O(N^{-6})$.

\subsubsection*{General argument}
To conclude we will do a more general analysis that does not rely on perturbation theory. As we will see in section \ref{sec:applications}, at large dimension the operator spectrum and the OPE coefficients of an extremal solution often asymptote to those of a generalized free solution. We show here that this implies that if we have solved the functional equations (numerically) with $N$ components, and $N$ is large enough so that the free asymptotics have kicked in, we can effectively complete the solution for arbitrarily high dimensional operators.

The full constraints of crossing symmetry on the extremal solution are written in the following way:
\bea
\hat \omega_n(0)+\sum_{m=0}^{M} a_{m} \hat \omega_n(\Delta_m)&=-\sum_{m=M+1}^{+\infty} a_{m} \hat \omega_n(\Delta_m),& \qquad 0&\leq n\leq N\\
\sum_{m=M+1} a_{m} \hat \omega_{N+n}(\Delta_m)&=-\hat \omega_{N+n}(0)- \sum_{m=0}^M a_{m} \hat \omega_{N+n}(\Delta),& \qquad 0&< n.
\eea
Truncating the solution means that we set to zero the righthand side of the first set of equations, while ignoring the constraints arising from the second set. We have seen that for extremal solutions we have $M=\frac{N-1}2$, so we make that choice, and use for the functional basis the one introduced in equation \reef{eq:basisfuncs}. 

Let us suppose that the exact solution has a spectrum which asymptotes to the generalized free boson for high enough scaling dimension, that is, the high energy spectrum consists of ``double-trace'' operators whose anomalous dimensions $\gamma_m\equiv \Delta_m-\Delta_m^B$ fall off with some power of $m$. For definiteness let us say that $\gamma_m$ is sufficiently small for $m>M^*$. The bounds derived in \cite{Mazac2019a} imply that the corresponding OPE coefficients must also approach the generalized free ones. In this case, as long as $M=N/2>M^*$, the righthand side of the first set equations is indeed small for all $n<N$. This follows from the asymptotic behaviour, determined in the same reference:
\bea
\beta_n(\Delta) &\underset{\Delta\to \infty} {\sim} \sin\left[\frac{\pi}2(\Delta-\Delta^B_n)\right]^2 \left(\frac{a_{\Delta_n}^{\text{free}}}{a_{\Delta}^{\text{free}}}\right) \frac{4 \Delta (\Delta_n^B)^2}{\Delta^4-(\Delta_n^B)^4},\qquad \text{with} \quad \Delta/\Delta_n\quad \text{fixed},\\
\beta_n(\Delta) &\underset{\Delta\to \infty} {\sim} \sin\left[\frac{\pi}2(\Delta-\Delta^B_n)\right]^2 \left(\frac{a_{\Delta_n}^{\text{free}}}{a_{\Delta}^{\text{free}}}\right) \frac{4 (\Delta_n^B)^2}{\Delta^3},\qquad \text{with} \quad \Delta_n\quad \text{fixed}, \label{eq:asymptotic}
\eea
with similar expressions for the $\alpha_n$. In particular the effect of tails on the functional equations for fixed $n$ falls off as $\sim\gamma_N^2/N^2$. In the OPE maximization problem, which will also be analysed numerically in the next section, $\gamma_n\sim n^{-2}$ which leads to an overall expected $1/N^6$ convergence rate, which matches what we found in perturbation theory. More generally, we expect a $1/N^2$ fall-off behaviour until $N$ is large enough that we reach the scale where anomalous dimensions are sufficiently small, followed then by the faster $1/N^6$ fall-off.

Let us now discuss the second set of equations which we ignored. Again for large enough $N$ that we reach the weakly coupled regime, these equations become
\bea
\label{eq:highdims}
a_{n}&=-\alpha_{n}(0)-\sum_{m=0}^{\frac{N-1}2} a_m \alpha_{n}(\Delta_m)+O(\gamma_N^2),& \qquad n&> \frac{N-1}2\\
a_n \gamma_n &=-\sum_{m=0}^{\frac{N-1}2} a_m \beta_n(\Delta_m)+O(\gamma_N^2),&\qquad n&> \frac{N-1}2
\eea
In fact, we may freely drop terms in the sums above for which $m$ is above the strongly coupled scale.
These equations tell us that at high scaling dimension the CFT data is determined entirely in terms of the strongly coupled region where operators have order one anomalous dimensions. 

Overall we see that using our proposed basis, the truncation procedure effectively converges as soon as $N$ is large enough that we reach the weakly coupled regime of small anomalous dimensions. This means that, in practice, we can use numeric algorithms to solve for the low energy spectrum, and then use equations \reef{eq:highdims} to complete the solution up to arbitrarily large scaling dimensions. The reason this is possible is because all functionals in the basis automatically encode the asymptotics of the generalized free spectrum. In particular any finite linear combination of functionals will have double zeros at the positions of the free boson spectrum for sufficiently high dimension, and so truncation preserves the asymptotics. Of course, it is clear that should the extremal solution asymptote to a generalized free fermion we would do well then to use the associated functional basis.

\section{Applications}
\label{sec:applications}
In this section we will test our proposed functional basis in concrete numerical applications. 

In the first application, we will bootstrap the generalized free boson solution using the fermionic functional basis. This is interesting for several reasons. Firstly, we have an exact solution to compare with. Secondly, the fermionic functionals are adapted to solutions to crossing that asymptote to a spectrum of the form $\Delta_n=1+2\Df+2n$, i.e. not the generalized free boson's $\Delta_n=2\Df+2n$. From the fermionic functionals perspective the free boson is as strongly coupled as it is possible to get, so many of our motivating arguments in the previous section aren't applicable. It is interesting then to see how our basis performs in this worse case scenario.

In our second application we will maximize the OPE coefficient of the first operator above the identity, whose dimension $\Delta_0$ we take to continuously interpolate between the bosonic value $2\Df$ and the fermionic value $1+2\Df$. We find that right up to $\Delta_0=1+2\Df$, where there is a discontinuity, the bound is saturated by a solution whose spectrum asymptotes to the generalized free boson's, and as such we will use the bosonic functional basis throughout. This is a concrete numerical test of the analytic arguments of the previous section. For $\Delta_0$ perturbatively close to $2\Df$ the extremal solution matches with what would be obtained by adding a $\phi^4$ interaction to a free scalar field in AdS$_2$, and as such we call this family of solutions the $\phi^4$ flow.

All computations in this section were performed with {\tt JuliBootS} \cite{Paulos:2014vya}. As an input, we need to provide to this program the functional actions $\alpha_n(\Delta),\beta_n(\Delta)$, analogously to $\omega^\partial(\Delta)\equiv\partial_z^{2n+1} z^{-2\Df}G_{\Delta}(z)|_{z=1/2}$ for the derivative basis. Unlike the latter, for the former we do not yet have a way of obtaining efficient rational function representations, and hence we simply computed a table of functional actions evaluated at several values of $\Delta$ and fed it into the linear programming algorithm. To improve the accuracy of the overall result, interpolation was then used to iteratively generate new points on the table in the vinicity of functional zeros until convergence was observed. Further details are available upon request.

Explicit forms of the functional kernels $f_\omega(z)$ for both the bosonic and the fermionic functionals for any external dimension $\Delta_{\phi}$ are given in appendix \ref{app:funcbasis}. We have chosen to focus on $\Delta_\phi=1/2$ for the fermionic basis and $\Delta_\phi=1$ for the bosonic one, where these $f_\omega$ take a simpler form. In this case we are able to compute the functional actions analytically, see appendix \ref{app:computation}, making the numerical evaluation of the tables faster. We emphasize that apart from this there is nothing special about these particular external dimensions: for a generic $\Delta_\phi$ it may take longer to compute the value of the functional actions, but once this is done the linear programming part of the algorithm proceeds in the same way. In particular, all nice properties of the functional bases, such as fast convergence, are generic.\footnote{As a sanity check, we have explicitly checked this up to $N=10$ for some non half-integer external dimension in the fermionic basis.}

\subsection{Bootstrapping bosons with fermions}
As we have reviewed in the previous section, in 1D CFTs the OPE maximization problem for an operator with dimension $2\Df$ is solved by the generalized free boson, with the exact functional providing this bound given by $\alpha_0^B$.\footnote{To be precise this assumes a certain minimum gap in the spectrum. This gap can be chosen to be the scaling dimension, always below $2\Df$, where $\alpha_0^B$ first becomes negative.} We now attempt to reproduce this result using the fermionic functional basis and compare with the same computation as done with the derivative basis. In detail the two basis used are:
\bea
\mathcal  W_N^\omega&\equiv \text{span}\left\{ \alpha^F_n, \beta^F_n, \qquad n=0,\ldots,\frac{N-2}2\right\},\\
\mathcal W_N^\partial&\equiv \text{span}\left\{ \partial_z^{1+2n}|_{z=\frac 12}, \qquad n=0,\ldots, N-1\right\}.
\eea

\begin{figure}%
\begin{center}
\includegraphics[width=15cm]{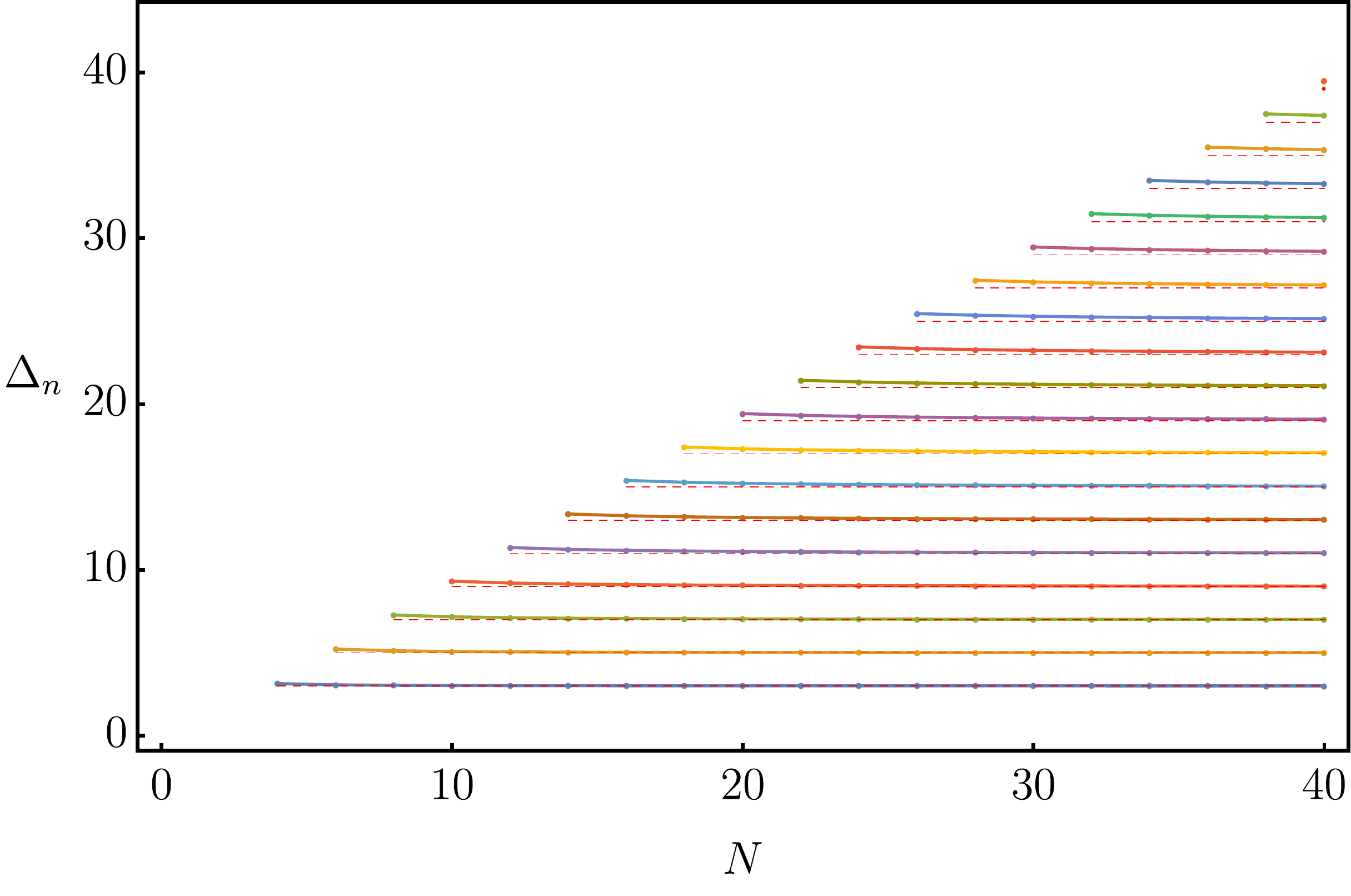}\\
\includegraphics[width=15cm]{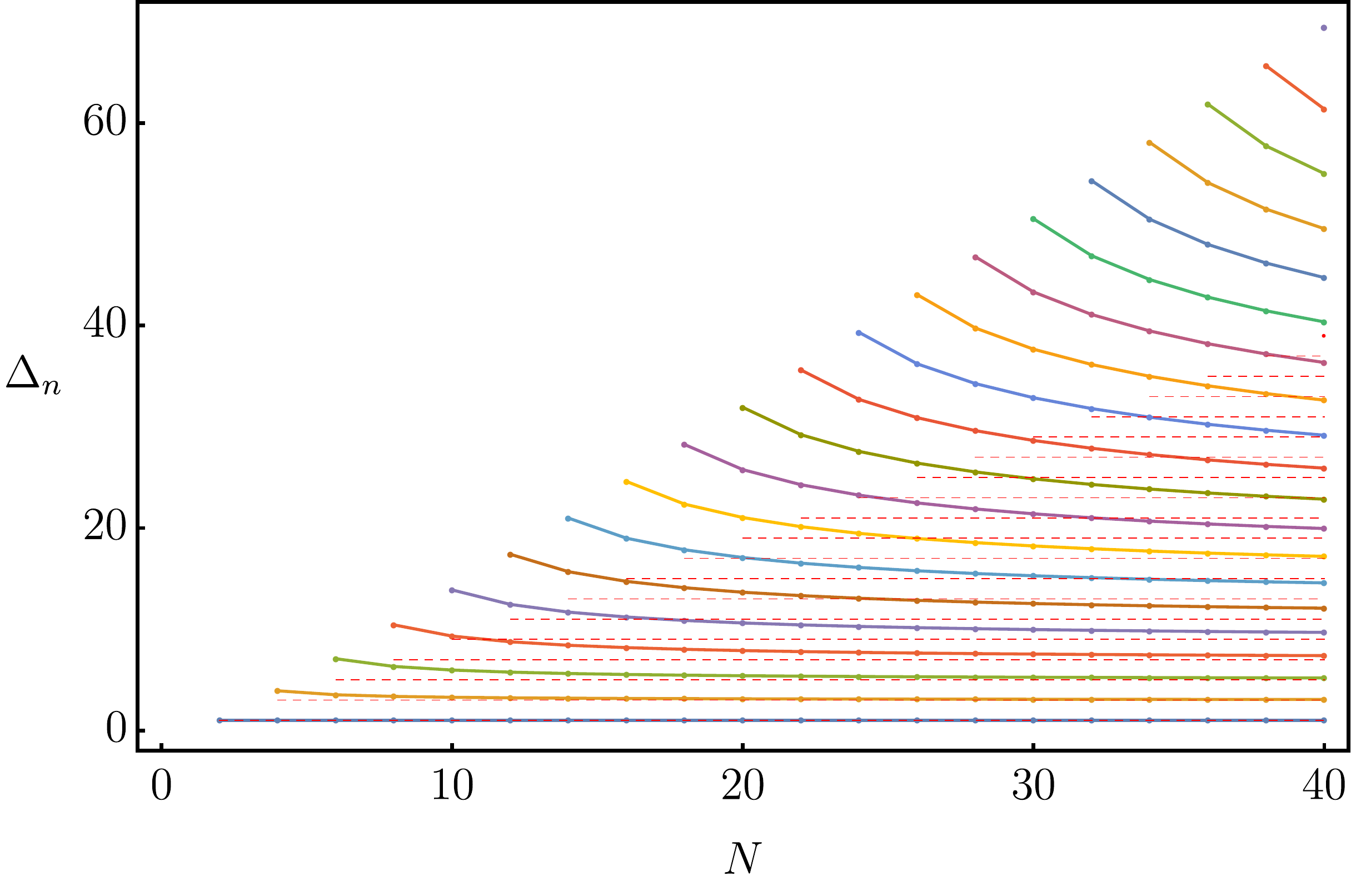}
\caption{Spectrum of the solution saturating an OPE maximization bound for $\Delta_0=2\Df$, with $\Df=1/2$, using the fermionic functional basis (top) and derivatives (bottom). As $N$ is increased by two units a new operator appears, while previous operators approach their correct values shown as dashed lines. With functionals the initial error is smaller and convergence is faster.}%
\label{fig:spectrum}%
\end{center}
\end{figure}

\begin{figure}%
\begin{center}
\begin{tabular}{cc}
\includegraphics[width=8cm]{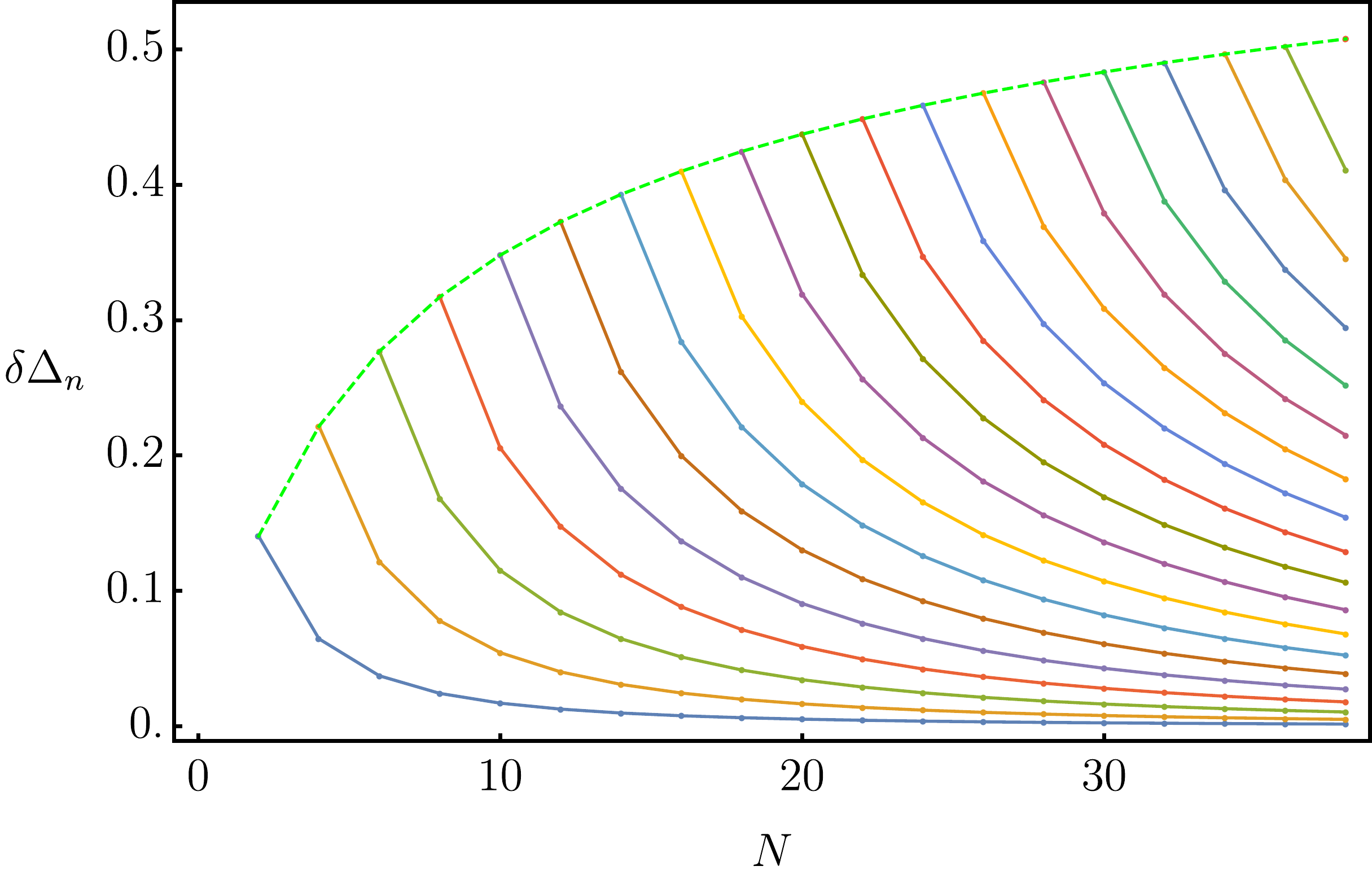}&
\includegraphics[width=8cm]{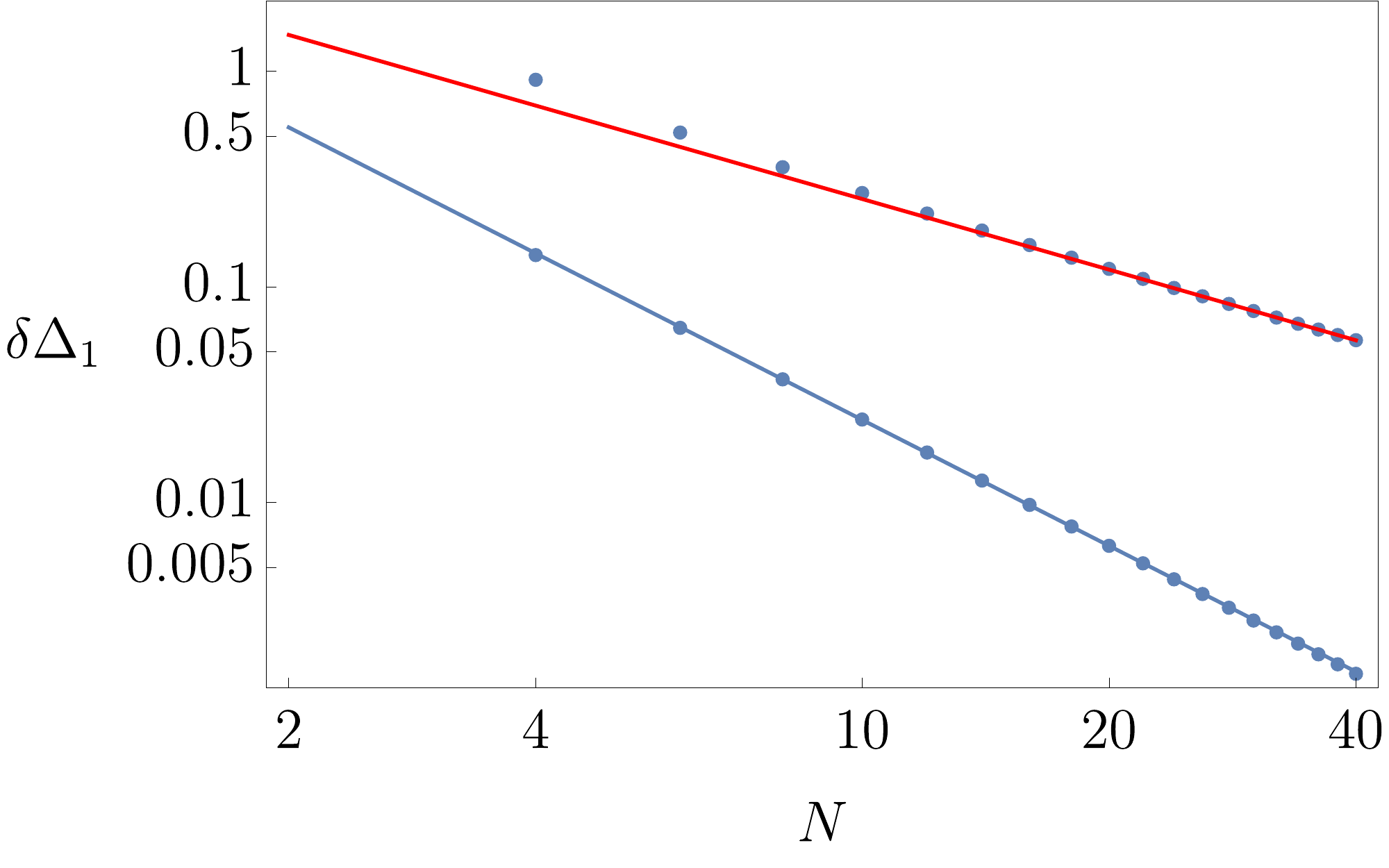}%
\end{tabular}
\caption{Spectrum of the solution saturating OPE maximization bound at $\Delta_0=2\Df$ with $\Df=1/2$. $\delta \Delta_n \equiv | \Delta_n- \Delta_n^B|$. Left: using the fermionic functional basis, as $N$ is increased by two units a new operator appears with initial bounded \emph{absolute} error in dimension, which converges to about $\sim 0.6$ at large $N$. As $N$ increases further this error then rapidly decreases. Right: Convergence of scaling dimension $\Delta_1$. In red the derivative basis, in blue the functional basis. The lines represent fits to a $1/N$ and $1/N^2$ behaviour respectively.}%
\label{fig:convergence}%
\end{center}
\end{figure}

We focus on $\Delta_\phi=1/2$, for which the functional actions can be computed more efficiently, as shown in appendix \ref{app:computation}. Our results are shown in figures \ref{fig:spectrum} and \ref{fig:convergence}. We find the functional basis outperforms the derivative basis in two different ways. Firstly, as expected for both bases the extremal solution gains a new operator every time that $N$ is increased by two units (meaning we have two additional functional components), however with functionals this operator appears already with a small initial error, which is in fact bounded in absolute value by $\sim 0.6$. No such bound is observed with the derivative basis, with this initial error seemingly growing linearly with $N$. After their initial appearance operators systematically approach their correct generalized free values as $N$ is increased further. In the derivative basis this takes longer than with functionals, since not only the initial error is larger, but also the rate at which the correct value is approached is worse. Experimentally we find convergence rates of order $N^{-2}$ and $N^{-1}$ for the functional and derivative bases respectively. The former is consistent with our analysis of the previous section setting $\gamma_n\sim 1$.

\begin{figure}
	\centering
	\includegraphics[width=10cm]{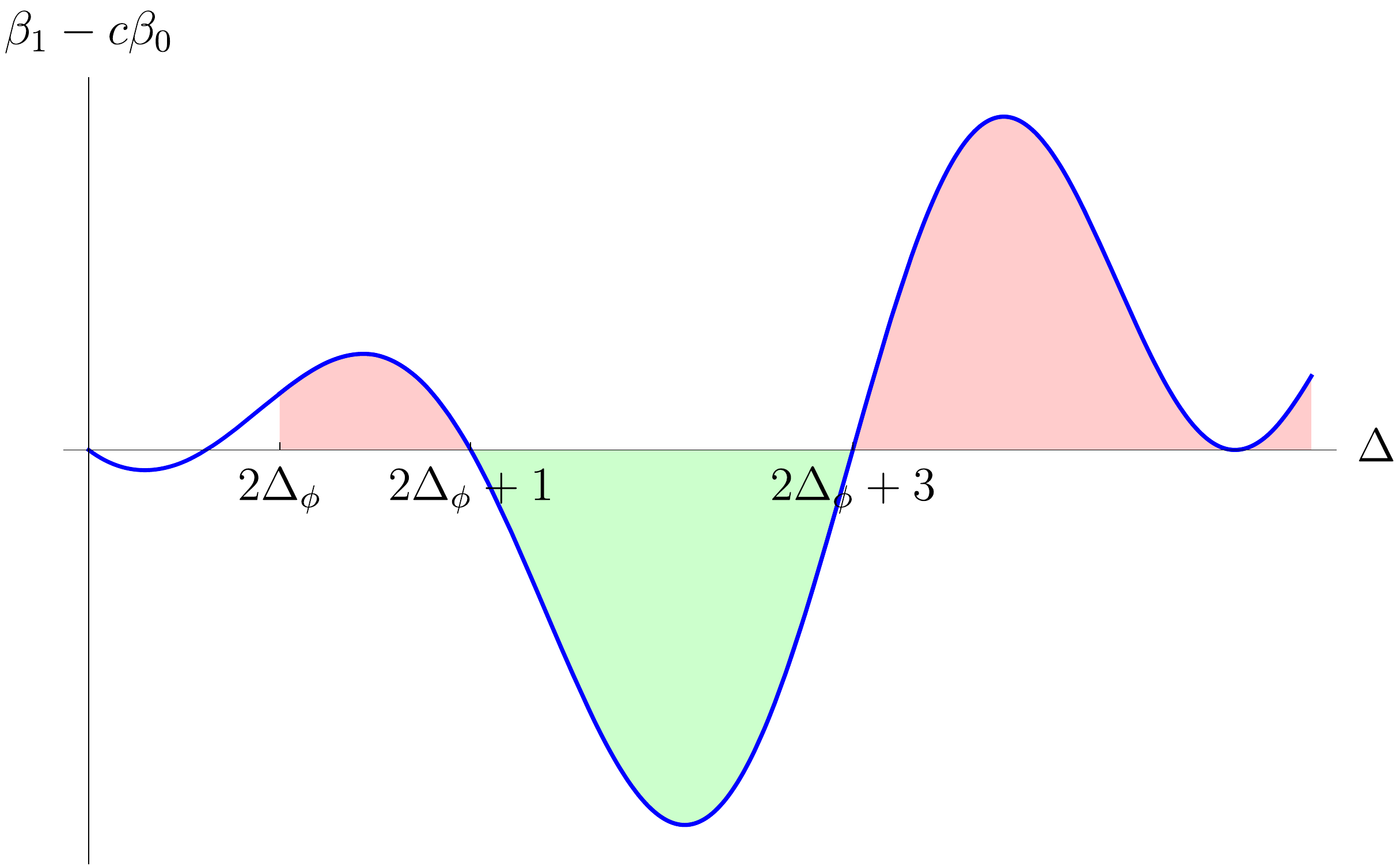}
	\caption{The single bin functional $\cup$, here shown for $\Df=\frac 12$. Contributions to the OPE in the region where the functional is negative must cancel those where it is positive. Given our gap assumptions on the extremal solution, this implies an operator must appear between $2\Df+1$ and $2\Df+3$.}
	\label{fig:binfunc}
\end{figure}
The reason why the initial error of operators is bounded when using the functional basis can be understood from the results of \cite{Mazac2019a}. There it was shown that by combining functionals it is possible to find lower bounds on the OPE density. This forces operators to appear in certain bins in scaling dimension space. These bounds can be obtained by combining a finite number of functionals, and hence hold even in our truncated numerical setting. To see this explicitly, in figure \ref{fig:binfunc} we plot the functional combination $\cup_1:=\beta_1^F-c\beta_0^F$, where $c$ is a suitably chosen coefficient. For $N\geq 4$, this functional lies inside $\mathcal W_N$. Any solution to crossing, and in particular the extremal solution for the same $N$, has to be compatible with the constraints imposed by this functional. This constraint is that contributions from the OPE density in regions where $\cup_1$ is positive have to be cancelled by those where it is negative. Since we are imposing a gap in the spectrum up to $2\Df$, the constraint from $\cup_1$ here implies a non-zero OPE density inside the region $(1+2\Df,3+2\Df)$. At extremality this will be achieved by a single operator inside this region, and so the error on the dimension of this operator is bounded by one unit (since the correct value is $2+2\Df$). This argument generalizes to higher values of $n$.

\subsection{OPE maximization, or $\phi^4$ flow}

For our next application, we will derive bounds on the OPE coefficient of an operator whose dimension $\Delta_0$ is varied between $2\Df$ and $2\Df+1$. As mentioned earlier, we will set $\Df=1$ for simplicity and use the bosonic functional basis since as it turns out, the extremal solution saturating the bound has a spectrum which rapidly asymptotes to that of a free boson. That is, we will now do the numerical version of what we examined in section \ref{sec:advantages}. The basis of functionals was given in \reef{eq:basisfuncs}.
Note that the existence of the $\alpha_0^{B,F}$ functionals imply that the exact bound will be saturated by the bosonic solution when $\Delta_0=2\Df$ and by the fermionic solution when $\Delta_0=1+2\Df$. Hence we expect that using the bosonic basis should be less optimal as we approach $\Delta_0=1+2\Df$ (but still better than derivatives, since in the worse case scenario we go back to a similar situation to that of the previous subsection). 

\begin{figure}%
	\begin{center}
		\includegraphics[width=12cm]{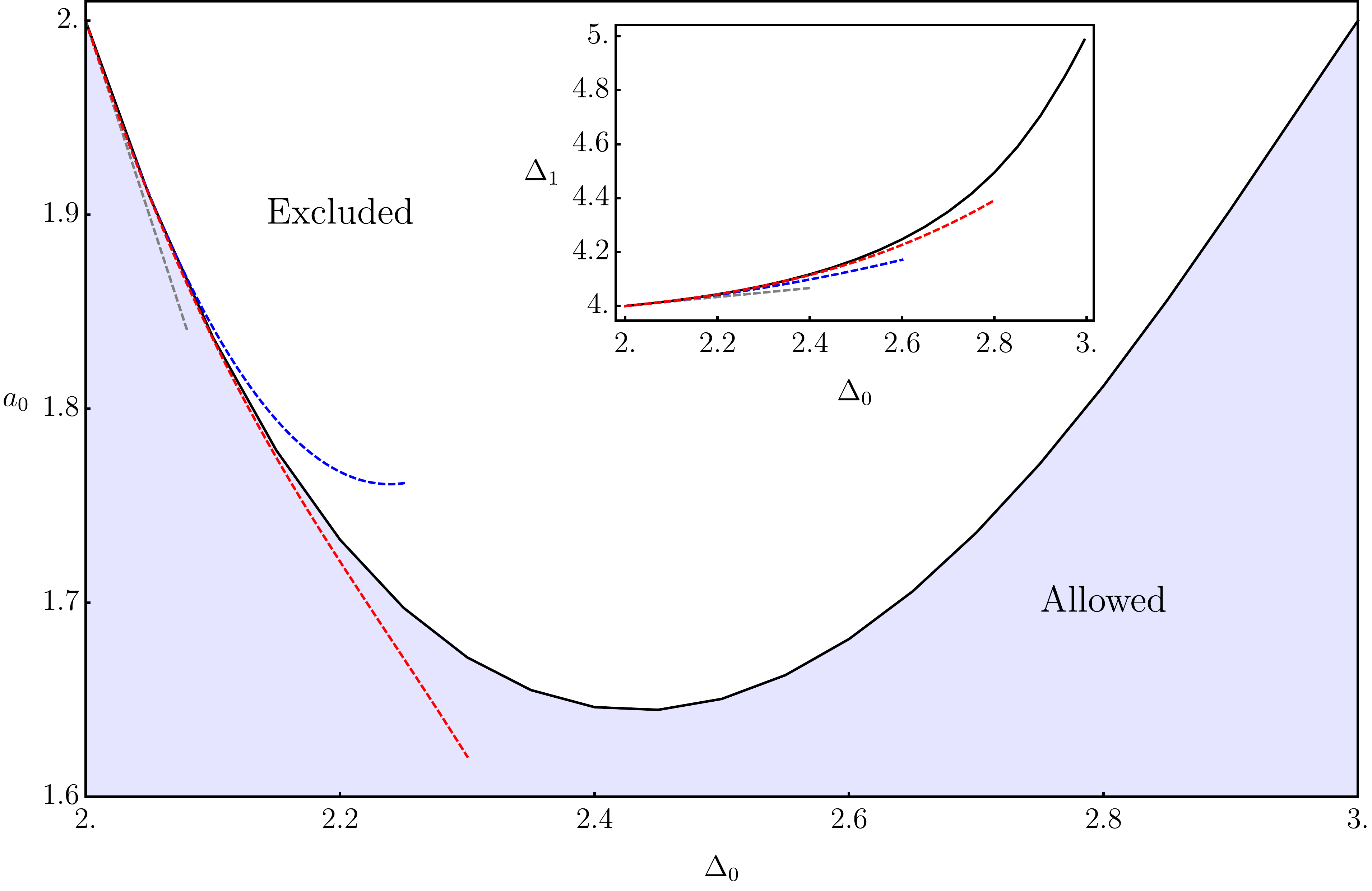}
		\caption{OPE maximization at $\Delta_{\phi}=1$ with $N=21$ components, although no visible changes are seen beyond $N=5$. Main plot: bound on the OPE coefficient of the first operator as a function of its dimension $\Delta_0$. Inset: dimension of the second operator $\Delta_1$ at extremality as a function of $\Delta_0$. In both plots, the dashed lines represent analytic perturbation theory computations up to cubic order in $g\equiv \Delta_1-2\Df$, as described in the main text.}
		\label{fig:opeflow1}%
	\end{center}
\end{figure}

\begin{figure}%
	\begin{center}
		\includegraphics[width=12cm]{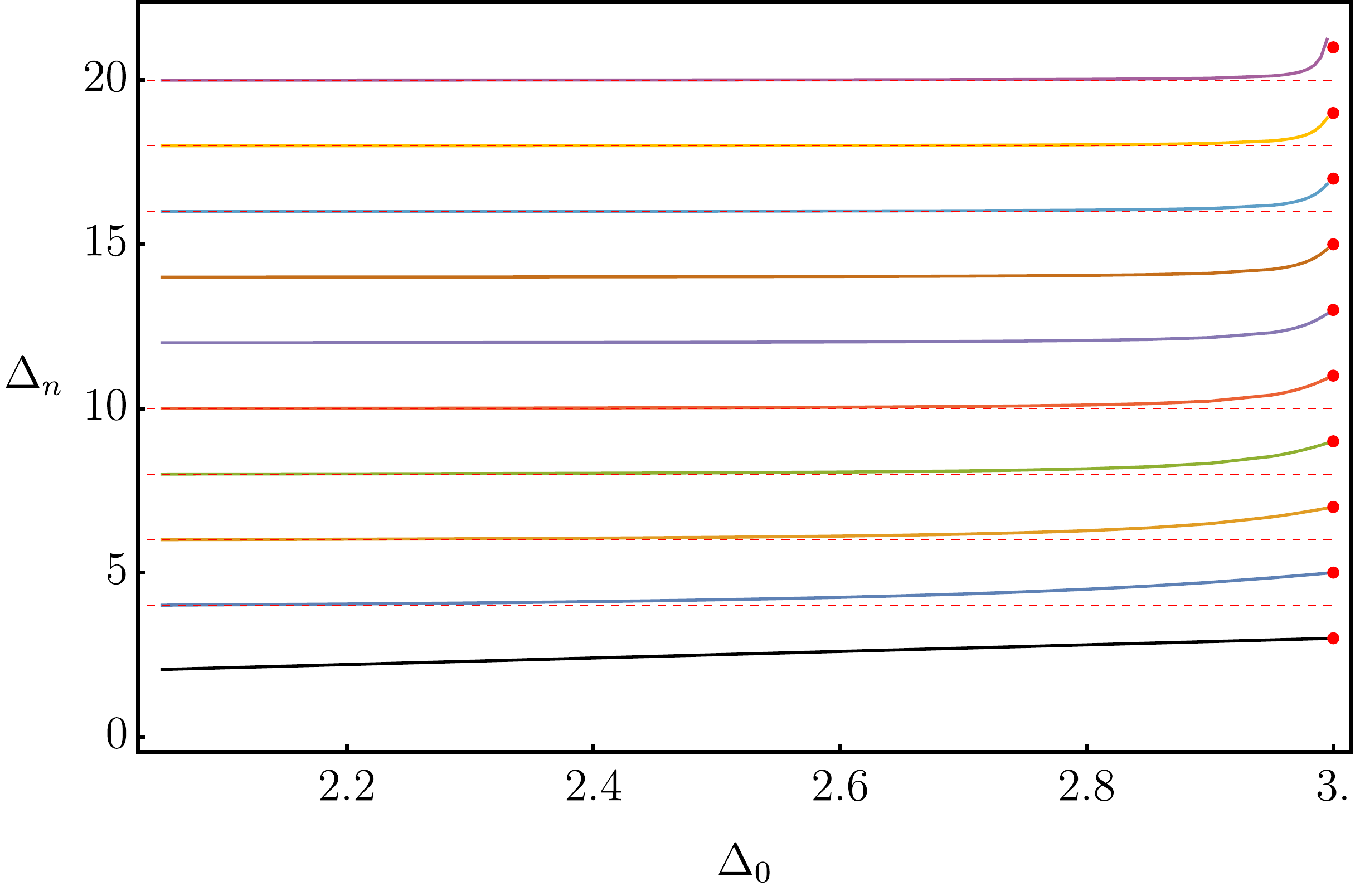}
		\caption{Extremal spectrum as a function of the lowest dimension operator with $N=21$ components. The dashed lines represent the generalized free boson spectrum, while the red dots stand for the free fermion spectrum.}
		\label{fig:opeflow2}%
	\end{center}
\end{figure}

Our results are summarized in figures \ref{fig:opeflow1} and \ref{fig:opeflow2}. In the main plot in figure \ref{fig:opeflow1} we show the upper bound on the OPE coefficient. The bound is essentially unchanged beyond truncations with $N=5$ with even $N=3$ already providing an excellent approximation to the bound except very close to $\Delta_0=1+2\Df=3$. The results shown have $N=21$. In the inset of the same figure we show the scaling dimension $\Delta_1$ of the subleading scaling operator (since $\Delta_0$ is fixed by hand) of the extremal solution saturating the bound.\footnote{A different way to think of this curve is as an upper bound on the dimension of the first subleading scalar in a solution to crossing, where the leading scalar is assumed to have dimension $\Delta_0$. In other words, the extremal solution that saturates the OPE bound can also be found by doing gap maximization on the subleading scalar.} The numerical results are in excellent agreement with the analytic results of reference \cite{Mazac2019a}, where CFT data for the $\phi^4$ deformation was determined up to two loops. Our numerics can be thought of as providing a non-perturbative completion of these results into the strongly coupled regime.

When $\Delta_0=2\Df$ or $\Delta_0=1+2\Df$ the OPE bound is saturated by the free values $2$ and $2\Df$ respectively (which for $\Df=1$ are the same), as expected, and $\Delta_1$ also correctly interpolates between $2\Df+2=4$ and $1+2\Df+2=5$. Inbetween those values, we have a non-trivial bound and an interacting solution to crossing. Figure \ref{fig:opeflow2} shows a more complete picture of the spectrum obtained from the same truncation with $N=21$. We see that most of the spectrum is essentially identical to that of the free boson for almost all values of $\Delta_0$, which explains why using the bosonic functional basis is such a good idea here. It is only when $\Delta_0$ approaches $1+2\Df$ that we see rapid rearrangements in the spectrum. For sufficiently large $\Delta$ the spectrum essentially jumps nearly discontinuously from that of the free boson to the free fermion one. This is consistent with the fact that there are no UV-preserving deformations of the generalized free fermion that do not introduce new states, as follows from the duality conditions \reef{eq:dualityfermion}. Presumably irrelevant contact interactions in AdS$_2$ should account for the spectrum close to the free fermion point.

\begin{figure}
	\centering
	\includegraphics[width=13cm]{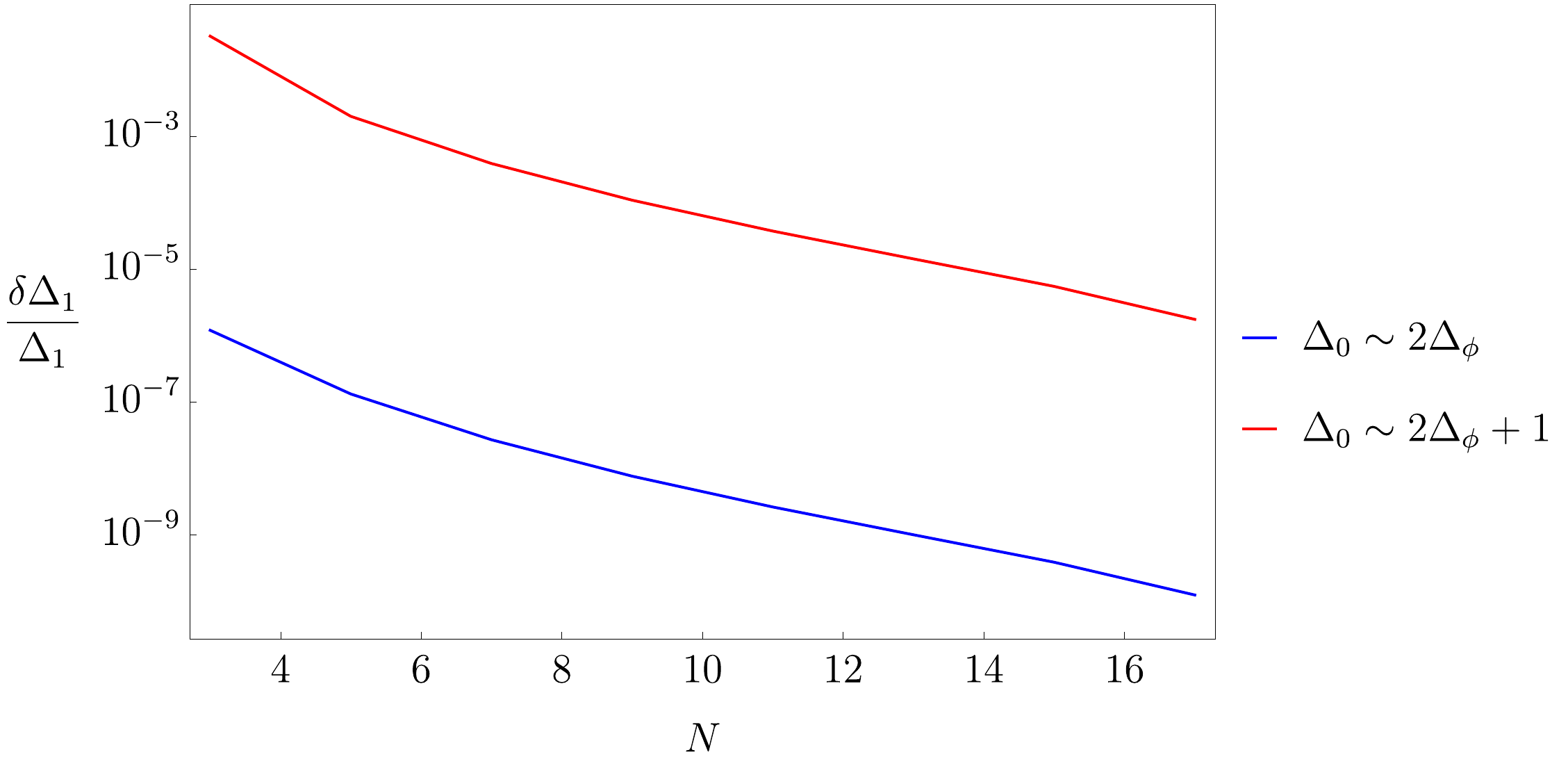}
	\caption{Convergence of OPE maximization. Shown are the relative errors in the determination of $\Delta_1$ (as compared to its value for $N=21$), for two distinct values of $\Delta_0$, namely close to the bosonic and fermionic points. Although the initial error is smaller in the former (since we are using the bosonic basis), the rate of convergence is the same in either case and compatible with a $1/N^6$ behaviour.}
	\label{fig:phi4convergence}
\end{figure}				

Finally, we can contrast these results with those using the derivative basis. In the latter, there is no special difference between setting $\Delta_0=2\Df$ or some other value, and so the convergence properties looks the same as those of the previous section. In particular, one again finds that operators come in with larger and larger initial errors, and converge with a power law falloff $\sim 1/N$. In figure \ref{fig:phi4convergence} we show convergence properties with the functional basis. Although it is hard to estimate the correct fall-off power from our data (we have considered only functionals up to $N=21$), the results are consistent with a very fast convergence rate $\sim N^{-6}$, as argued in section \ref{sec:advantages}.

We observe that as we increase $N$, new operators appear with a value of the scaling dimension which is essentially already their correct final value. That is, after an operator appears, increasing $N$ does not significantly change its OPE or scaling dimension. In this sense, we find quite remarkably that the solution to the numerical problem {\em converges}. This is in line with the expectations of section \ref{sec:advantages}: once we have successfully bootstrapped the region of the spectrum where scaling dimensions are very different from the free ones, there is very little interest in pursuing the computation, since the OPE data of higher dimension operators may be determined by equations \reef{eq:highdims}. We have checked that those equations are indeed satisfied to high accuracy.

\section{Outlook - higher dimensions}
The main conclusion of this note is that the subset of constraints  arising from the $z=\bar z$ section of the crossing equation are efficiently captured by the functional bases introduced in \cite{Mazac2019a}. In particular, we have argued and shown that for 1D CFTs these functionals provide what is perhaps the best possible choice of basis with which to compute numerical bounds. 
Our results make even more exciting the possibility that there are analytic functionals in higher dimensions, analogous to the ones in 1D, which would allow us to distinguish between different spin channels. If such functionals can be constructed, even discounting their analytic implications, they will surely revolutionize what is possible to do in the numerical bootstrap. Since such functionals are not yet available, we would like to comment on what can be done right away using the functionals we do have. 

Upon restriction to the $z=\bar z$ line, higher dimension solutions to crossing equation take the form.
\bea
\sum_{\Delta,\ell} a_{\Delta,\ell} F_{\Delta,\ell}(z,z|\Df)=0.
\eea
Hence there are now many kinds of functional actions, depending on spin:
\bea
\omega(\Delta)\to \omega(\Delta,\ell).
\eea
However, it is straightforward to show that the functional actions still encode generalized free asymptotics in every spin channel, i.e. we have relations such as
\bea
\alpha_n^B(\Delta_m^B,\ell)=\delta_{n,m},\qquad \mbox{for}\quad m\geq n
\eea
independently of spin.\footnote{This is a consequence that the derivation of the functional action \reef{eq:funcaction}, with the characteristic sine squared oscillations, actually depends on very little details of the conformal blocks.}

When bootstrapping higher dimensional CFTs, the class of functionals typically used amounts to derivatives in the two cross-ratios $z,\bar z$, which may be taken along directions parallel and transverse to the $z=\bar z$ line. So, one obvious thing to do is to replace purely parallel derivatives by the functional basis proposed here. More generally one could imagine Taylor expanding the $F_{\Delta,\ell}(z,\bar z)$ along the $z=\bar z$ line and acting with our functionals on each coefficient function. However, there are good reasons to believe that by simply adding our functionals to the existing approach will already lead to significant and in some cases dramatic improvements.

The first basic reason is that even the set of $z=\bar z$ constraints in higher dimensions are non-trivial, especially for moderate and large values of $\Delta_\phi$. Consider for instance the typical bootstrap problem, which is a bound on the leading scalar dimension, while allowing for higher spin operators with dimensions above or equal to the unitarity bound $d-2+\ell$. Since all higher $D$ CFTs are also $D=1$ CFTs, this bound must be equal to or stronger than the 1D result $1+2\Df$ for $1+2\Df\leq d$.\footnote{When we set a gap from a 1D perspective we are gapping all spin channels in higher dimensions. The restriction then arises since we don't want to gap spin 2 operators above the unitarity bound.}
The point now is that even mustering sufficient computation firepower to obtain a bound as strong as the $D=1$ result takes some work. As a simple example for instance, in $D=4$ (cf. \cite{Poland2012a}, figure 2) with $\Df=3/2$, the usual derivative bound only becomes competitive with the $D=1$ result after using at least 21 derivative components, whereas the same result would be obtained with a single functional ($\beta_0^F$).

More interestingly, higher dimensional bounds can in fact sometimes be saturated by 1D solutions, where our functionals can then provide the optimal bound immediately. A trivial example is that if we take the free theory solution in any spacetime dimension, from a 1D perspective this is nothing but the generalized free boson, for which we already have associated extremal functionals. In particular, this predicts that the free theory solution maximizes the OPE bound on the scalar operator of dimension $2\Df=d-2$. Hence, in the neighbourhood of the free points, using the functional basis above should not be too bad, since using a truncation containing a single functional $\alpha_0^B$ (i.e. the analog of using a single derivative) already yields the exact optimal bound at that point.

\begin{figure}%
\begin{center}
\includegraphics[width=10cm]{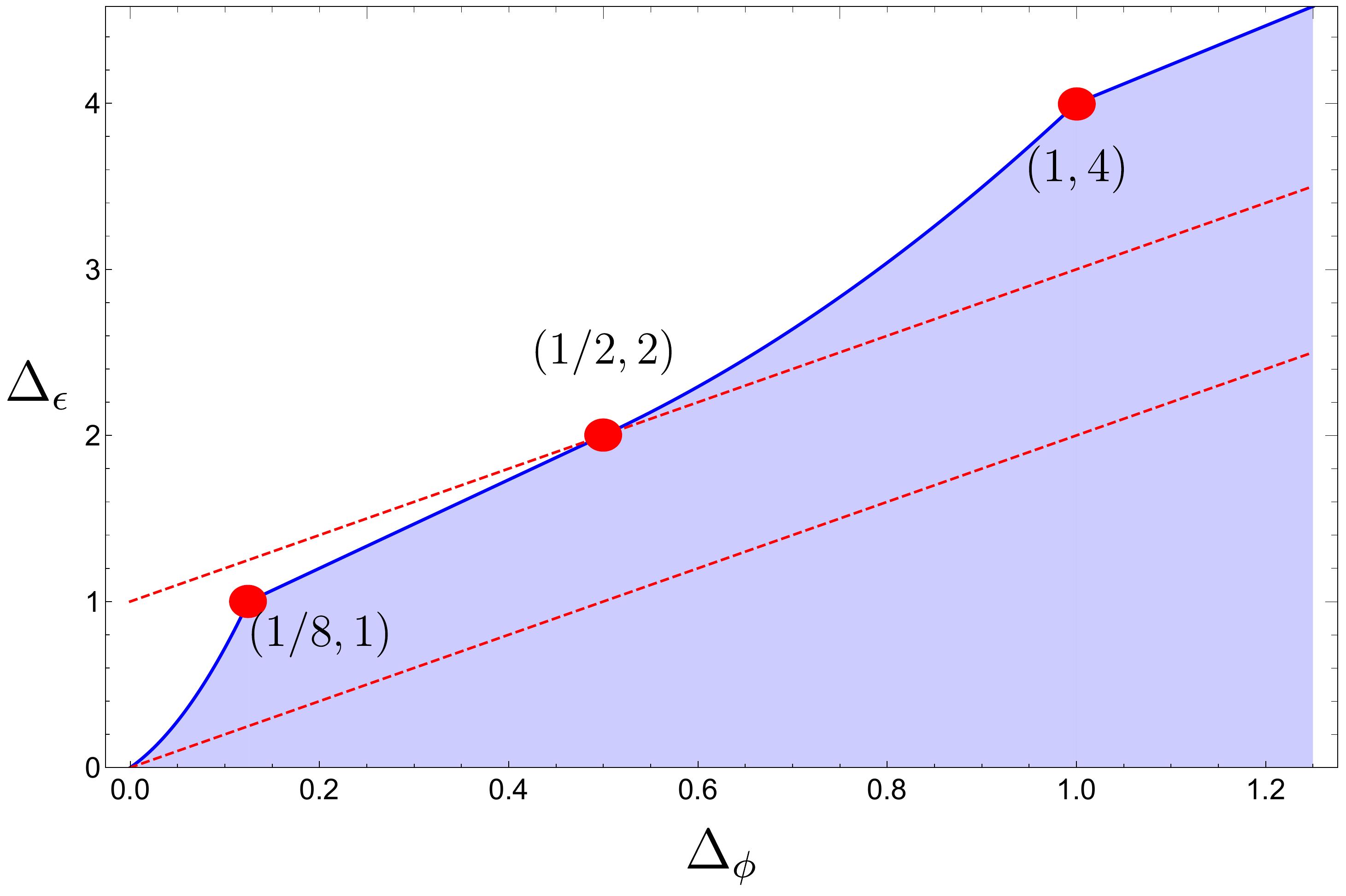}%
\caption{Schematic bound on leading scalar dimension $\Delta_\epsilon$ in 2D CFTs. Three notable points $(\Delta_\phi,\Delta_\epsilon)$ are the 2D Ising model $\langle \sigma \sigma \sigma \sigma\rangle$ correlator at $(1/8,1)$, the large $m$ limit of the minimal model $\phi_{1,2}$ four-point correlator at $(1/2,2)$ and the 2D Ising $\langle \epsilon \epsilon \epsilon \epsilon\rangle$ correlator at $(1,4)$. The dashed lines correspond to the 1D generalized free fermion ($\Delta_\epsilon=1+2\Df$) and boson ($\Delta_{\epsilon}=2\Df$) theories.}
\label{fig:2dbound}%
\end{center}
\end{figure}

Two more interesting examples occur in $D=2$. There the bound on the leading scalar dimension looks schematically as in figure \ref{fig:2dbound}.\footnote{The first 2D bound appeared in \cite{Rychkov:2009ij}, and a wider plot where the second kink at (1,4) is visible has appeared recently in \cite{Gowdigere:2018lxz}. However, the existence of this kink was communicated to us by S. El-Showk a long time ago, who seems to have been the first to notice it.}
The first example is at the point $(\Delta_\phi, \Delta_\varepsilon)=(1/2,2)$, where the bound crosses the 1D generalized free fermion line. This point corresponds to the $m\to \infty$ limit of the minimal model correlator of four $\phi_{1,2}$ \cite{Liendo2013},\footnote{This is also the correlator of four energy operator in the 4-state Potts model at criticality, see for example \cite{Gorbenko:2018dtm}.} which for $z=\bar z$ is indeed the same as the generalized free fermion. Hence, at this point we already have the extremal functional available which determines the exact optimal bound, namely $\beta_0^F$. Establishing this with the derivative basis, even approximately, is highly non-trivial. The bound determined by $\beta_0^F$ alone is suboptimal everywhere below $\Df=1/2$ as it should, and above this value it gaps the spin-2 channel as well, so there is no inconsistency with higher-D results.

The second example is at $(\Delta_\phi, \Delta_\varepsilon)=(1,4)$ which is the four $\varepsilon$ correlator for the 2D Ising model. This correlator restricted to the $z=\bar z$ line becomes a generalized free boson. From the 1D perspective the operator of dimension $2\Df=2$ is the 2D stress tensor, and $\varepsilon$ is the operator with dimension $2+2\Df=4$. In particular the OPE bound on an operator of dimension $2\Df$ is saturated by the free boson, which in higher $D$ terms translates into the statement that central-charge minimization will be saturated by the $\varepsilon$ four point function, with corresponding functional $\alpha_0^B$.

In both cases, we see that in our basis truncations with even a single component already yield exact optimal bounds, so that transverse derivatives are only required to disentangle the spin structure of the extremal spectrum. It is reasonable to expect that in the vicinity of these points including our functionals in the basis should lead to great improvements in current numerical results. It would be very interesting to investigate this in a more systematic fashion.

\vspace{1 cm}
\section*{Acknowledgments}
We would like to acknowledge A. Kaviraj, D. Maz\'a\v{c}, S. Rychkov, A. Vichi and especially E. Trevisani for discussions and comments at various points of this collaboration. 
BZ is supported by the National Centre of Competence in Research SwissMAP funded by the Swiss National Science Foundation.

\appendix

\section{Functional kernels}
\label{app:funcbasis}
In this section we show how to obtain the explicit kernels corresponding to functionals $\alpha_n, \beta_n$. For further details the reader should consult \cite{Mazac2019a}. We start with the results:
\begin{subequations}
\ba
f_{\beta_0^F}(z|\Df)=
-\kappa(\Df)
&\frac{2z-1}{w^{3/2}}\left[\, _3\tilde{F}_2\left(-\frac{1}{2},\frac{3}{2},2
   \Df+\frac{3}{2};\Df+1,\Df+2;-\frac{1}{4 w}\right)+\right.\\
	&\,\;\;\left.+\frac{9}{16 w} \,
   _3\tilde{F}_2\left(\frac{1}{2},\frac{5}{2},2 \Df+\frac{5}{2};\Df+2,\Df+3;-\frac{1}{4w}\right)\right],
\label{eq:allaf}
\ea
\ba
f_{\alpha_0^F}(z|\Df)=
\kappa(\Df)\frac{2(z-2)(z+1)}{(2z-1)w^{3/2}}
&\left[
{}_3\tilde{F}_2\left(-\frac{1}{2},-\frac{1}{2},2\Df +\frac{3}{2};\Df +2,\Df +2;-\frac{1}{4 w}\right)+\right.\\
+\frac{(2 \Df +3) (2 \Df +5)}{16 w} &{}_3\tilde{F}_2\left(\frac{1}{2},\frac{1}{2},2 \Df +\frac{5}{2};\Df +3,\Df +3;-\frac{1}{4 w}\right)-\\
-\frac{3 (4 \Df +5)}{256 w^2}&\left.{}_3\tilde{F}_2\left(\frac{3}{2},\frac{3}{2},2 \Df +\frac{7}{2};\Df +4,\Df +4;-\frac{1}{4 w}\right)\right].
\label{eq:fAlphaTilde}
\ea
\end{subequations}
where $w\equiv z(z-1)$ and we have added an extra argument to the kernel $f(z)\to f(z|\Df)$ to keep track of which $\Df$ it corresponds to. Here ${}_3\tilde F_2$ stands for the regularized hypergeometric function, $w=z(z-1)$ and the normalization factor is given by
\be
\kappa(\Df) = \frac{\Gamma(4\Df+4)}{2^{8\Df+5}\Gamma(\Df+1)^2}\,.
\ee
Recall that $g_\omega(z)=- (1-z)^{2\Df-2} f_{\omega}(\frac{1}{1-z})$ in the fermionic case (for bosons, the overall minus sign isn't there).
We can now constructed shifted functionals:
\bea
f_{s\omega_{m}^F}(z|\Df)=\left[\frac{1+z(z-1)}{z(z-1)}\right]^{2m} f_{\omega_0^F}(z|\Df+m)
\eea
where $\omega$ can be $\alpha$ or $\beta$. The corresponding shifted functionals $s\beta_m^F$, $s\alpha_m^F$ satisfy the correct duality conditions for $\Delta_n>\Delta_m$, and hence we have e.g.
\bea
\beta_n^F=\sum_{m=0}^{n} q_n\,  s\alpha_n^F+r_n\, s\beta_n^F
\eea
for some $q_n$, $r_n$. The constants may be determined by imposing the full duality relations \reef{eq:dualityfermion}. We note that the solution to this orthonormalisation step is not currently known in closed form for all $n$, i.e. it has to be done case by case, except for special values of $\Df$. However, it is clear that for the purpose of the numerical computations we are interested in any basis will do, and so we may as well work with the shifted functionals instead of the orthonormal ones.

Now let us turn to the bosonic functionals. We now define:
\bea
f_{s\omega^B_{m+1}}(z|\Df)=-\frac{1}{z(z-1)} \left[\frac{1+z(z-1)}{z(z-1)}\right]^{2m} f_{\omega_0^F}(z|\Df+3/2+m).
\eea
That is, the bosonic kernels can be obtained from the fermionic ones. The kernels above correspond to shifted bosonic functionals analogous to the fermionic ones, and again an orthonormalisation procedure may be applied if so wished, by imposing the duality conditions \reef{eq:dualityboson}. The above should be used for $m\geq 0$. The $\alpha_0^B$ functional must be treated separately, and is given by:
\bea
f_{\alpha_0^B}(z|\Df)=-z(z-1)\left[f_{\alpha_0^F}(z|\Df-3/2)+\frac{1}{2\Df-1} f_{\beta_0^F}(z|\Df-3/2)\right].
\eea

\section{Computation of the functional action for special cases}
\label{app:computation}
In this section we compute the action of the bosonic functionals for external dimension $\Delta_\phi=1$ and the fermionic ones for $\Delta_\phi=1/2$. We will give the explicit forms of these functionals and show how to evaluate them. The techniques shown here apply in the bosonic case with integer external dimension and fermionic case with half integer external dimension, where the functional kernels have a simpler form.

\subsection{The functional kernels}

The functionals act on $F_{\Delta}(z)$ via the formula:

\bea
\omega_n[F_{\Delta}]\equiv \omega_n(\Delta)=\frac 12 \int_{\frac 12}^{\frac 12+i\infty}\!\!\ud z f_{\omega_n}(z) F_\Delta(z)+\int_{\frac 12}^1 \ud z g_{\omega_n} (z) F_\Delta(z) \label{eq:contourapp}
\eea
where $g(z)=\pm (1-z)^{2\Df-2}f\left(\frac 1{1-z}\right)$ and the +(-) sign for the boson (fermion) basis respectively. This can be massaged into the alternative expression:
\begin{equation}
\omega_n(\Delta)=[1\pm \cos\pi(\Delta-2\Df)]\int_0^1 \ud z\, g_{\omega_n}(z)\frac{G_{\Delta}(z)}{z^{2\Df}} \label{eq:wint}.
\end{equation}
For the fermionic functionals at external dimension $\Delta_\phi=1/2$, the functional kernels are
\begin{equation}
\begin{aligned}
g^-_{\beta_n}(z)&=\frac{2 \Gamma(2+2n)^2}{\pi^2 \Gamma(3+4n)}\left[\frac 1z P_{1+2n}\left(\frac{2-z}z\right)-P_{1+2n}(2z-1)\right],\\
g^-_{\alpha_n}(z)&=\frac{2\, \Gamma(2+2n)^2}{\pi^2 \Gamma(3+4n)}\left[\frac 1z DP_{1+2n}\left(\frac{2-z}z\right)-DP_{1+2n}(2z-1)-\right.\nonumber 
\left. \frac{\Gamma(2+2n)^2}{\Gamma(4+4n)}\,\frac{G_{\Delta_n}(1-z)}{1-z}\right]=\\
&=\frac 12 \partial_n g_{\beta_n^{-}}(z)-\frac{2}{\pi^2}\frac{\Gamma(2+2n)^4}{\Gamma(3+4n)\Gamma(4+4n)}\,\frac{G_{\Delta_n}(1-z)}{1-z}\,.
\end{aligned}
\end{equation}
where $\Delta_n=2n+2$, $P_{n}(z)$ is the Legendre polynomial and the derivative with respect to its parameter is:
\begin{equation}
\begin{gathered} \label{eq:DP}
DP_n(y):=P_{n}(y)\log\left(\frac{1+y}2\right)+2\sum_{k=0}^{n-1}\left[(-1)^{k+n}\frac{2k+1}{(n-k)(n+k+1)}\,P_{k}(y)\right]=\\
=\frac{\Gamma(2n+1)}{\Gamma(n+1)^2} \partial_n\left[\frac{\Gamma(n+1)^2}{\Gamma(2n+1)}\, P_n(y) \right]\,.
\end{gathered}
\end{equation}
The motivated reader can check these expressions are perfectly compatible with those of appendix \ref{app:funcbasis}.
The bosonic functional kernels for external dimension are $\Delta_\phi=1$ are instead
\begin{equation}
\begin{gathered}
g^+_{\beta_{m}}(z)=\frac{2}{\pi^2}\frac{\Gamma(2+2m)^2}{\Gamma(3+4m)}\left[P_{2m+1}\left(\frac{2-z}{z}\right)+P_{2m+1}\left(2z-1\right)\right.\\\left.-P_{1}\left(\frac{2-z}{z}\right)-P_{1}\left(2z-1\right)\right]\,,\\
g^+_{\alpha_m}(z)=\frac 12\partial_m g^+_{\beta_m}(z)-\frac{2}{\pi^2}\frac{\Gamma(2+2m)^4}{\Gamma(3+4m)\Gamma(4+4m)}\, G_{\Delta_m}(1-z)\,.
\end{gathered}
\end{equation}
Numerically, the integral \eqref{eq:wint} converges only for $\Delta>\Delta_n$, and below that value we must use \reef{eq:contourapp}. Instead, we will now compute \eqref{eq:wint} exactly for $\Delta>\Delta_n$ and analytically continue the result to general $\Delta$. 

\subsection{Computation of beta functionals}
Let us start by showing how to compute the integral for the $\beta$ functionals. First of all we notice that $P_{1+2n}(\frac{2}{z}-1)$ is an eigenfunction of the Casimir $\calC_2=z^2(1-z)\frac{d^2}{dz^2}-z^2 \frac{d}{dz}$ with eigenvalue $c_2(\Delta_n)=\Delta_n(\Delta_n-1)=(2n+2)(2n+1)$. We use the fact that the measure $z^{-2}$ makes the Casimir operator self adjoint
\begin{equation}
\int_0^1 \frac{dz}{z^2}f(z) (\calC_2 g(z))=\left[(1-z)(f(z)g'(z)-f'(z)g(z)) \right]_0^1+\int_0^1 \frac{dz}{z^2} (\calC_2f(z))  g(z)\,.
\end{equation}
Combining this with the fact that both $G_\Delta(z)$ and $P_{2n+1}(\frac{2}{z}-1)$ are eigenfunctions of the Casimir, we get
\begin{equation}
\begin{gathered}
\int_0^1 \frac{dz}{z^2} P_{1+2n} \left( \frac{2}{z}-1\right) G_\Delta(z)=\frac{1}{c_2(\Delta)-c_2(\Delta_n)} \frac{\Gamma (2 \Delta )}{\Gamma (\Delta )^2}
\end{gathered} \label{eq:PCas}\,.
\end{equation}
The l.h.s. converges for $\Delta>\Delta_n$ only, but we use this formula as the analytic continuation. There is a pole at $\Delta=\Delta_n$, which when combined with the prefactor in \eqref{eq:wint} is responsible for the single zero of ${\beta_n}$, see \eqref{eq:dualityfermion}.

There is no such trick for the second part of the $\beta$ functional. Evaluating the Legendre polynomial for a given $n$ just leaves us to evaluate a finite number of integrals of the kind
\begin{equation}
I_a(\Delta)=\int_{0}^1 \frac{dz}{z^2}z^a G_\Delta(z)\,.
\end{equation}
This can be evaluated for generic values of $a$ and turns out to be
\begin{equation}
I_a(\Delta)=\frac{\, _3F_2(\Delta ,\Delta ,a+\Delta -1;2 \Delta ,a+\Delta ;1)}{a+\Delta -1}\,.
\end{equation}
However, it can be convenient to compute $I_a(\Delta)$ recursively, since this is numerically faster. Using the fact that the conformal block is an eigenvalue of the Casimir operator
\begin{equation}
\begin{gathered}
0=\int_{0}^1 \frac{dz}{z^2}z^a \left( \calC_2-c_2(\Delta)\right)  G_\Delta(z)= \\
=\frac{\Gamma (2 \Delta )}{\Gamma (\Delta )^2}
-\int_{0}^1 \frac{dz}{z^2} \calC_2 (z^a) G_\Delta(z) -c_2(\Delta)I_a(\Delta)=\\=\frac{\Gamma (2 \Delta )}{\Gamma (\Delta )^2}+(c_2(a)-c_2(\Delta))I_a(\Delta)-a^2 I_{a+1}(\Delta)\,,
\end{gathered}
\end{equation}
we get the recursion relation
\begin{equation}
I_{a+1}(\Delta)=\frac{1}{a^2}\frac{\Gamma (2 \Delta )}{\Gamma (\Delta )^2}+\frac{c_2(a)-c_2(\Delta)}{a^2} I_a(\Delta) \label{eq:recursionI}\,.
\end{equation}
Now, we can find the value of $I_a(\Delta)$ for any $a \in \mathbb{Z}$ using that
\begin{align}
I_0(\Delta)&=\frac{\Gamma (2 \Delta )}{(\Delta -1) \Delta \Gamma (\Delta )^2}\,, \\
I_1(\Delta)&= \frac{\Gamma (2 \Delta ) \Gamma (\Delta +1) \left(\psi ^{(1)}\left(\frac{\Delta }{2}\right)-\psi ^{(1)}\left(\frac{\Delta +1}{2}\right)\right)}{2 \Gamma (\Delta )^3} \,,
\end{align}
where for the second one we used equation (B.11) of \cite{2010arXiv1011.4546M}. By using this recursion relation there is no need to evaluate $_3F_2$ at unit argument, and this helps speeding up the numerics.

\subsection{Computation of the alpha functionals}
The computation of the $\alpha$ functional is a bit more complicated. Using the derivative relation \eqref{eq:DP} between $DP_n$ and $P_n$  and the result \eqref{eq:PCas} we get that

\begin{equation}
\begin{gathered}
\int_0^1 \frac{dz}{z^2} DP_{2n+1}\left(\frac{2}{z}-1 \right) G_\Delta(z)=\\=\left[2\left( H_{2n+1}-H_{4n+2}\right) +\frac{2\Delta_n-1}{c_2(\Delta)-c_2(\Delta_n)}  \right] \frac{1}{c_2(\Delta)-c_2(\Delta_n)} \frac{\Gamma (2 \Delta )}{\Gamma (\Delta )^2}\,,
\end{gathered}
\end{equation}
where $H_{n}$ is the n$^\text{th}$ Harmonic number. We can see that this integral has a double pole at $\Delta=\Delta_n$, and therefore ${\alpha_n}$ is finite at this point, see \eqref{eq:dualityfermion}.

The term with $G_{\Delta_n}(1-z)$ can be computed in terms of $_4F_3$ with unit argument by using the result (3.38) of \cite{Liu:2018jhs}, which we report here for completeness:
\begin{multline}
		\int_0^1 \frac{dz}{z^2}\left( \frac{z}{1-z}\right) ^p G_\Delta(z) G_{\Delta_n}(1-z)=\\
	=\frac{\Gamma \left(2 \Delta _n\right) \Gamma (p+\Delta -1)^2 \Gamma \left(-p-\Delta +\Delta _n+1\right) }{\Gamma \left(\Delta _n\right){}^2 \Gamma \left(p+\Delta +\Delta _n-1\right)}\\\times \, _4F_3\left(\Delta ,\Delta ,p+\Delta -1,p+\Delta -1;2 \Delta ,p+\Delta -\Delta _n,p+\Delta +\Delta _n-1;1\right)+\\+\frac{\Gamma (2 \Delta ) \Gamma \left(p+\Delta -\Delta _n-1\right) \Gamma \left(-p+\Delta _n+1\right){}^2 }{\Gamma (\Delta )^2 \Gamma \left(-p+\Delta +\Delta _n+1\right)}\\\times \, _4F_3\left(\Delta _n,\Delta _n,-p+\Delta _n+1,-p+\Delta _n+1;2 \Delta _n,-p-\Delta +\Delta _n+2,-p+\Delta +\Delta _n+1;1\right)\,.
\end{multline}
Finally, we are left with expanding the $DP_{2n+1}(2z-1)$ term and doing a finite number of integrals of the kind
\begin{equation}
J_a(\Delta)=\int_0^1 \frac{dz}{z^2}z^a \log (z) G_\Delta(z)\,,
\end{equation}
which have the solution
\begin{equation}
J_a(\Delta)=-\frac{\, _4F_3(\Delta ,\Delta ,a+\Delta -1,a+\Delta -1;2 \Delta ,a+\Delta ,a+\Delta ;1)}{(a+\Delta -1)^2}\,.
\end{equation}
When computing the action of the functionals the bottleneck is given by the evaluation of the $_4F_3$ at unit argument, and so we would like to compute the least possible amount of them. By noting that $J_a(\Delta)=\partial_a I_a(\Delta)$ and deriving formula \eqref{eq:recursionI}, we get
\begin{equation}
J_{a+1}(\Delta)=-\frac{2}{a^3} \frac{\Gamma (2 \Delta )}{\Gamma (\Delta )^2}+\frac{a+2 c_2(\Delta )}{a^3}I_a(\Delta) +\frac{c_2(a)-c_2(\Delta )}{a^2}J_a(\Delta)
\end{equation}
so that every $J_a(\Delta)$ can be expressed in terms of $J_1(\Delta)$ and $J_0(\Delta)=-\frac{\Gamma (2 \Delta )}{(\Delta -1)^2 \Gamma (\Delta +1)^2}$.

\small
\parskip=-10pt
\bibliography{biblio_numerics}
\bibliographystyle{utphys}

\end{document}